\documentclass[10pt]{article}
\usepackage{graphicx} 
\usepackage{amsmath}
\usepackage{authblk}
\usepackage{amssymb}
\usepackage{amsthm}
\usepackage{booktabs}
\usepackage{longtable}
\usepackage[round,authoryear]{natbib}

\usepackage[
    colorlinks=true,
    citecolor=blue,
    linkcolor=blue,
    urlcolor=blue
]{hyperref}

\usepackage[a4paper, left=1.1in, right=1.1in, top=1in, bottom=1.2in, footskip=.7in]{geometry}

\usepackage{enumitem}

\newlist{assumptions}{enumerate}{1}
\setlist[assumptions]{
  label=\textbf{(A\arabic*)},
  ref=A\arabic*,
  leftmargin=*,
  itemsep=0.5em
}

\newtheorem{theorem}{Theorem}
\newtheorem{proposition}{Proposition}

\newtheorem{definition}{Definition}
\newtheorem{lemma}{Lemma}

\newtheorem{corollary}{Corollary}

\theoremstyle{remark}
\newtheorem{remark}{Remark}

\title{Gaussian Process Prediction under Gaussian Input Error:
A First-Order Criterion for Local Improvement}

\author[1]{Rezoanoor Rahman\thanks{Corresponding author. 
Email: \texttt{rrahman1@isrt.ac.bd}}}
\author[2]{Fariha Taskin}

\affil[1]{Department of Statistics, The Ohio State University}
\affil[2]{Department of Mathematical Sciences, 
The University of Texas at Dallas}

\date{}

\begin{document}

\maketitle

\begin{abstract}
We study when Gaussian input error reduces prediction mean squared error (MSE) in Gaussian process regression, where inputs may represent spatial locations, time points, or vectors of covariate values. Our analysis concerns kriging adjusted for location error (KALE), developed by
\citet{cressie2003loc_error} as an optimal linear unbiased predictor under the corresponding second-order model. \citet{cervone2015loc_error} showed that KALE prediction MSE can be smaller than in the error-free input setting for a single training observation under a squared-exponential covariance and independent Gaussian input errors, without additive observation noise. With the covariance parameters and nominal inputs held fixed, we derive an exact first-order criterion for local improvement for arbitrary finite training sample sizes, allowing additive observation noise. The criterion applies to sufficiently smooth, stationary, isotropic covariance functions. In addition, for the squared-exponential covariance, we obtain sufficient conditions expressed through distances among the nominal training inputs and the prediction target. Finally, for the one-dimensional exponential covariance, we establish local improvement for every finite set of distinct nominal training inputs and every target outside that set.

\end{abstract}



\section{Introduction}


Gaussian process (GP) models provide a flexible framework for regression and prediction by specifying a mean function and a covariance function that describes dependence among observations. GP models appear throughout the applied and methodological literature: emulating and calibrating expensive computer and climate models \citep{computer_exp_Gp, bayesian_calibration_computer}, mapping disease risk in epidemiology \citep{diggle_disease_mapping}, and guiding black-box optimization \citep{gp_ucb}. In this article, we use the generic term ``input'' to refer to the index at which the GP is evaluated. In spatial statistics, the input is a geographic location, typically represented by two-dimensional coordinates \citep{cressie_spatial_book, stein1999}. In time series analysis, the input is time \citep{roberts2013}. In machine learning applications, the input is generally a vector of feature values; structured inputs, such as images, may also be represented as vectors, for example by flattening the corresponding pixel arrays \citep{lee2018}.

A common assumption in GP models is that the inputs are known without error \citep{gp_ml_book}.  In practice, this assumption is often violated in applications where GPs are widely used. For example, wildlife telemetry tags report animal locations with GPS error \citep{gps_error}, household coordinates released with nationally representative health surveys are deliberately displaced for confidentiality \citep{burgert2013geographic}, and tissue coordinates in spatial genomics assays are registered across slides only approximately \citep{jones2023alignment}. More generally, in spatial modeling \citep{barber2006modelling} and time series analysis \citep{girard_input_error}, the inputs attached to observations may themselves be measured with error.

Input errors in GP models alter the covariance among observations and their cross-covariance with the prediction target. Hence, treating input errors simply as additive measurement error in the response may fail to capture the induced covariance structure and can lead to misspecified predictors \citep{cervone2015loc_error}. To account for this uncertainty, \citet{cressie2003loc_error} introduced kriging adjusted for location error (KALE) for spatial prediction, which incorporates location uncertainty through the first two moments induced by the input errors. Under this second-order framework, KALE is the best linear unbiased predictor (BLUP) and does not require the observations to be jointly Gaussian. Although we use the term ``input'' throughout this article to refer more generally to the indices of a GP, we retain the name KALE for consistency with the existing literature.

Beyond the KALE framework, several notable studies on input uncertainty appear in the literature. For example, \citet{diggle_spatial_error} studied spatial prediction in the presence of positional error and demonstrated, using simulations and environmental pollution data, that accounting for positional uncertainty can alter predictive distributions and substantive conclusions. Within Gaussian process regression, \citet{girard_input_error} propagated input uncertainty through multi-step-ahead forecasts, whereas \citet{mchutchon_input_error} used a local linear approximation to represent Gaussian input noise as an additional input-dependent contribution to predictive uncertainty.

Measurement error in predictors can limit prediction accuracy even when its distribution is incorporated into the prediction model. For example, \citet{spicker2025prediction_error} established a higher lower bound on validation MSE under additive predictor measurement error than in the corresponding error-free setting. They also examined the resulting loss of accuracy in feed-forward neural networks and linear regression. Related effects have been reported for binary outcomes. In probit regression, random measurement error can increase the Brier score and reduce the area under the receiver operating characteristic curve \citep{khudyakov2015measurement_error}. In logistic regression, measurement error and differences in predictor measurement across development and validation settings can affect calibration, discrimination, and overall predictive accuracy \citep{rosella2012measurement_error, luijken2019measurement_error}.

In contrast, \citet{cervone2015loc_error} established a counterintuitive result for GP regression. Under a squared-exponential covariance function and independent and identically distributed (iid) Gaussian input errors, they derived a necessary and sufficient distance-based condition under which KALE can achieve a lower prediction mean squared error (MSE) than predictions based on observations obtained at the exact nominal inputs. However, their result is limited to a single training observation and does not consider cases with additive measurement errors.

In this article, we extend the single-observation result of \citet{cervone2015loc_error} to arbitrary finite training sample sizes. We first derive an exact first-order criterion for local improvement in KALE prediction MSE under stationary, isotropic covariance functions that are twice continuously differentiable near the relevant lags, with iid Gaussian input errors and with or without additive observation noise. With the covariance parameters, observation-noise variance, nominal inputs, and prediction target held fixed, this criterion determines the direction of the local MSE change when the first-order coefficient is nonzero. For the squared-exponential (SE) covariance, we derive distance-based sufficient conditions for local improvement. For the one-dimensional exponential covariance, we show that local improvement holds at every unobserved target when the nominal training inputs are pairwise distinct. Finally, we illustrate these results through forecasting and missing-value imputation in sparsely sampled time series. For this example, we express the SE sufficient conditions in terms of regular-grid spacing and target location, and examine the corresponding changes in prediction MSE numerically. 

\section{Preliminaries}

\subsection{Notation}

Throughout this article, matrices are written in bold uppercase and vectors in bold lowercase. The exceptions are inputs $s,t\in\mathbb R^p$ and random vectors such as the input errors $U_i\in\mathbb R^p$ and Gaussian perturbations $Z\in\mathbb R^p$, which are written in plain type. For a positive integer $n$, we use $[n]$ to denote the set of the first $n$ positive integers, $\{1,2,\ldots,n\}$, and $\mathbf I_n$ is the $n\times n$ identity matrix. For a matrix $\mathbf M$, $[\mathbf M]_{ij}$ is its $(i,j)$th entry; for a vector $\mathbf v$, $[\mathbf v]_k$ is its $k$th entry. For a symmetric matrix $\mathbf A$, $\lambda_{\min}(\mathbf A)$ and $\lambda_{\max}(\mathbf A)$ are its smallest and largest eigenvalues, respectively, and $\mathbf A\succeq 0$ and $\mathbf A\succ0$ mean that $\mathbf A$ is positive semidefinite and positive definite. We write $\|\mathbf v\|_2$ and $\|\mathbf M\|_F$ to denote the Euclidean norm of $\mathbf v$ and the Frobenius norm of $\mathbf M$, respectively. $N_p(\boldsymbol\mu,\boldsymbol\Sigma)$ is the $p$-variate normal distribution, $\Phi$ is the standard normal distribution function, and iid abbreviates independent and identically distributed. Finally, for functions $g_1$ and $g_2$, with $g_2$ nonzero near the stated limit, we write $g_1=o(g_2)$ if $g_1/g_2\to0$ and $g_1=O(g_2)$ if $|g_1/g_2|$ remains bounded as that limit is approached.

\subsection{Gaussian processes with input error}\label{sec:model}

Let $f$ be a zero-mean Gaussian process (GP) on $\mathbb R^p$ with stationary, isotropic covariance function
\begin{equation*}
c(s,t)=\operatorname{Cov}\{f(s),f(t)\}
=c_{\boldsymbol\theta}(\|s-t\|_2),
\qquad s,t\in\mathbb R^p.
\end{equation*}
Here, $\boldsymbol\theta=(\tau^2,\ell)$ contains the process variance $\tau^2=c_{\boldsymbol\theta}(0)$ and the lengthscale $\ell$ \citep{cressie_spatial_book,gp_ml_book}. In the presence of input error, the $i$th observation is recorded at a nominal input $s_i$, but the actual input is $s_i+U_i$. Then the corresponding observation is
\begin{equation}
\label{eq:observation_model}
y_i=f(s_i+U_i)+\epsilon_i,\qquad i\in[n].
\end{equation}
We assume the input errors $U_i$ are independent and identically distributed (iid) with mean zero and have a known distribution. The observation errors $\epsilon_i$ are iid with mean zero and variance $\sigma_{\epsilon}^2 = \eta\tau^2$, where $\eta=\sigma_\epsilon^2/\tau^2\ge0$ is the noise-to-signal variance ratio. The latent process $f$, the input-error sequence $\{U_i\}$, and the observation-error sequence $\{\epsilon_i\}$ are mutually independent \citep{cressie2003loc_error,barber2006modelling}. We index observations by $i$ rather than by $s_i$ since $y_i$ depends on $i$ through $U_i$, and two observations may share a nominal input.

We distinguish the following two settings and keep them separate throughout this research.
\begin{enumerate}
\item \emph{Observation noise-free setting}: $\eta=0$ and $y_i=f(s_i+U_i)$.
\item \emph{Noisy setting}: $\eta>0$, and the response also contains additive observation noise.
\end{enumerate}
Both settings use the same form of the KALE predictor. However, the conditions ensuring non-singularity of the observation covariance matrix, and hence a well-defined predictor, differ between the two settings.

\subsection{Kriging adjusted for input error}\label{sec:kale}

Suppose our prediction target is the latent value $f(s^*)$ at an exactly known input $s^*$. Under input error, the joint distribution of $\mathbf y_n=(y_1,\dots,y_n)^\top$ is generally not Gaussian. When the observation errors are Gaussian, the observation vector is Gaussian conditional on the input errors $U_1,\dots,U_n$. Thus, its marginal distribution is a mixture of Gaussians, and the conditional mean $\mathbb E\{f(s^*)\mid\mathbf y_n\}$ is generally nonlinear in $\mathbf y_n$ with no closed form. \citet{cressie2003loc_error} therefore proposed kriging adjusted for location error (KALE), which uses only the first two moments of $\mathbf y_n$ and $f(s^*)$. Define $\mathbf K(\sigma_u^2)$ as the covariance matrix of the latent process evaluated at the perturbed training inputs, and $\mathbf k^*(\sigma_u^2)$ as the cross-covariance vector between the observations and the target, with elements
\begin{equation*}
[\mathbf K(\sigma_u^2)]_{ij}
=\mathbb E[\,c(s_i+U_i,s_j+U_j)],
\qquad\text{and}\qquad
[\mathbf k^*(\sigma_u^2)]_i
=\mathbb E[\,c(s^*,s_i+U_i)].
\end{equation*}
The expectations are over the input errors, whose distribution is indexed by its variance $\sigma_u^2$. Since $f$ has mean zero and is independent of both error sequences, the observation covariance matrix is $\operatorname{Var}(\mathbf y_n)=\mathbf K_\eta(\sigma_u^2)=\mathbf K(\sigma_u^2)+\eta\tau^2\mathbf I_n$, and $\operatorname{Cov}\{\mathbf y_n,f(s^*)\}=\mathbf k^*(\sigma_u^2)$. Then the KALE predictor is
\begin{equation}
\label{eq: kale}
\hat f(s^*)
=\mathbf k^*(\sigma_u^2)^\top
\mathbf K_\eta(\sigma_u^2)^{-1}\mathbf y_n,
\end{equation}
and its corresponding MSE is
\begin{equation}
\label{eq: kale_MSE}
M_{n,\eta}(\sigma_u^2)
=\tau^2-\mathbf k^*(\sigma_u^2)^\top
\mathbf K_\eta(\sigma_u^2)^{-1}\mathbf k^*(\sigma_u^2).
\end{equation}
Because the mean is known, $\hat f(s^*)$ is the best linear unbiased predictor of $f(s^*)$ \citep{cressie2003loc_error}, and $M_{n,\eta}(\sigma_u^2)$ is its exact MSE. Neither equation~\eqref{eq: kale} nor equation~\eqref{eq: kale_MSE} requires $\mathbf y_n$ or the $\epsilon_i$ to be Gaussian. At $\sigma_u^2=0$, $\hat f(s^*)$ reduces to kriging based on exact inputs. As a result, $M_{n,\eta}(0)$ is the zero-input-error baseline, evaluated at the same observation-noise variance. In the noise-free setting, we write $\mathbf K=\mathbf K_0$ and $M_n=M_{n,0}$. For the covariance functions considered here, distinct nominal inputs ensure that the covariance matrix is invertible at $\sigma_u^2=0$ when $\eta=0$.

\subsection{Closed forms under Gaussian input error}\label{sec:closed}

For a general error distribution, closed forms of $\mathbf K$ and $\mathbf k^*$ may not be available, in which case the required expectations can be approximated by Monte Carlo integration \citep{cressie2003loc_error}. Under Gaussian input error, they reduce to a Gaussian smoothing of the covariance function. This gives closed forms for the two covariances we study and makes the dependence on $\sigma_u^2$ explicit.

Let $U_i\overset{\mathrm{iid}}{\sim}N_p(\mathbf 0,\sigma_u^2\mathbf I_p)$. For $\sigma^2\ge0$, define the smoothed covariance
\begin{equation}
\label{eq: cov_GP_smooth}
c_{\sigma^2}(s,t)=\mathbb E[\,c(s+Z,t)],
\end{equation}
where $Z\sim N_p(\mathbf 0,\sigma^2\mathbf I_p)$. For $\sigma^2=0$, we get the original covariance with no input error. Since $c$ is stationary and $U_i-U_j\sim N_p(\mathbf 0,2\sigma_u^2\mathbf I_p)$ for $i\ne j$,
\begin{equation}
\label{eq: genenral_expectation}
[\mathbf K(\sigma_u^2)]_{ij}
=
\begin{cases}
c_{2\sigma_u^2}(s_i,s_j), & i\ne j,\\
\tau^2, & i=j,
\end{cases}
\qquad\text{and}\qquad
[\mathbf k^*(\sigma_u^2)]_i=c_{\sigma_u^2}(s^*,s_i).
\end{equation}
The target--observation cross-covariances are smoothed at variance $\sigma_u^2$, whereas the off-diagonal observation covariances are smoothed at variance $2\sigma_u^2$.

\begin{definition}[SE-GE and Exp-GE models]
\label{def:models}
Consider the observation model \eqref{eq:observation_model} with iid Gaussian input errors, $U_i\overset{\mathrm{iid}}{\sim}N_p(\mathbf 0,\sigma_u^2\mathbf I_p)$. We consider two covariance choices for the latent GP. The \emph{SE-GE model} that uses the squared-exponential (SE) covariance:
\begin{equation*}
c(s,t)
=\tau^2\exp\left\{
-\frac{\|s-t\|_2^2}{\ell^2}
\right\},
\qquad s,t\in\mathbb R^p,
\end{equation*}
and the \emph{Exp-GE model} that uses the exponential covariance for one-dimensional inputs, $p=1$,
\begin{equation*}
c(s,t)
=\tau^2\exp\left\{
-\frac{|s-t|}{\ell}
\right\},
\qquad s,t\in\mathbb R.
\end{equation*}
Here, SE and Exp identify the latent covariance function, while GE indicates Gaussian input error. In both models, $\ell$ denotes the lengthscale parameter. Each model includes settings without additive observation noise ($\eta=0$) and with additive observation noise ($\eta>0$).
\end{definition}

The SE covariance function used in this article differs from the $\exp\{-r^2/(2\ell^2)\}$ convention of \citet{gp_ml_book}: our $\ell$ equals $\sqrt2$ times theirs, and all subsequent thresholds use this convention. With $r=\|s-t\|_2$, the smoothed covariances can be written as
\begin{equation*}
\text{SE-GE:}\quad
c_{\sigma^2}(s,t)
=\tau^2\Bigl(1+\frac{2\sigma^2}{\ell^2}\Bigr)^{-p/2}
\exp\Bigl(-\frac{r^2}{\ell^2+2\sigma^2}\Bigr),
\qquad \sigma^2\ge0,
\end{equation*}
and
\begin{equation*}
\text{Exp-GE:}\quad
c_{\sigma^2}(s,t)
=\tau^2e^{\sigma^2/(2\ell^2)}
\Bigl\{
e^{-r/\ell}\Phi\Bigl(\frac{r}{\sqrt{\sigma^2}}
-\frac{\sqrt{\sigma^2}}{\ell}\Bigr)
+e^{r/\ell}\Phi\Bigl(-\frac{r}{\sqrt{\sigma^2}}
-\frac{\sqrt{\sigma^2}}{\ell}\Bigr)
\Bigr\},
\qquad \sigma^2>0.
\end{equation*}
At $\sigma^2=0$, the Exp-GE smoothed covariance is defined as the original exponential covariance. The SE-GE expression follows from the moment generating function of a noncentral $\chi^2$ distribution, and the Exp-GE expression follows by splitting the integral at zero and completing the square. Detailed derivations of these closed-form expressions are provided in Appendix~\ref{derivation_closed_form}.

For the theoretical results in the following sections, we assume the covariance parameters and observation-noise variance are fixed to isolate the effect of input uncertainty on prediction. Estimating these parameters introduces additional uncertainty \citep{zimmerman_cressie_1992}, whose theoretical treatment is beyond the scope of this article.

\section{Main result}\label{sec:exact}

We consider a sufficiently smooth stationary, isotropic covariance function, both with and without additive observation noise. Throughout, we assume $U_i\overset{\mathrm{iid}}{\sim}N_p(\mathbf 0,\sigma_u^2\mathbf I_p)$; and the matrices $\mathbf K(\sigma_u^2)$ and $\mathbf k^*(\sigma_u^2)$ are obtained from the smoothed covariance in equation~\eqref{eq: genenral_expectation}. This smoothing averages the covariance around the original lag. For small input-error variance, the change in covariance depends on the local curvature of the covariance function. In this section, we combine these covariance changes to determine their effect on prediction MSE.

For a covariance function $c(s,t)$, let $\nabla^2c(s,t)$ denote its Laplacian with respect to $s$:
\begin{equation*}
\nabla^2c(s,t)
=
\sum_{a=1}^p
\frac{\partial^2c(s,t)}{\partial s_a^2}.
\end{equation*}
A prime at zero denotes the right derivative with respect to $\sigma_u^2$ at $\sigma_u^2=0$, taken entrywise for vectors and matrices. To determine the sign of $M_{n,\eta}'(0)$, we first impose the following assumptions.
\begin{assumptions}[
label=\textnormal{(G\arabic*)}, ref=G\arabic*
]
\item \label{ass:smoothness_general}
The covariance function, expressed as a function of the lag, is twice continuously differentiable in a neighborhood of every target lag $s^*-s_i$, $i\in[n]$, and every off-diagonal training lag $s_i-s_j$, $i\ne j$.
\item \label{ass:pd_general}
The observation covariance matrix at zero input-error variance is positive definite:
\begin{equation*}
\mathbf{K}_\eta(0)
=
\mathbf{K}(0)+\eta\tau^2\mathbf{I}_n
\succ0.
\end{equation*}
\end{assumptions}
Assumption~(\ref{ass:smoothness_general}) allows us to use a second-order Taylor expansion at the relevant lags which holds for the SE covariance and for Mat\'ern covariance functions with smoothness parameter $\nu>1$ \citep{gp_ml_book}. For $0<\nu\le1$, Mat\'ern covariance functions, including the exponential covariance, are twice continuously differentiable if the lags are nonzero. They therefore satisfy this assumption when the nominal training inputs are pairwise distinct and the target differs from every training input. On the other hand, assumption~(\ref{ass:pd_general}) ensures that the kriging weights and the KALE prediction MSE are well defined in a neighborhood of zero input-error variance. This assumption holds whenever the additive observation noise has positive variance. Without additive observation noise, it holds when the nominal training inputs are pairwise distinct and the covariance function is strictly positive definite.

Under assumption~(\ref{ass:smoothness_general}), using a second-order Taylor expansion at each relevant lag, followed by averaging over the Gaussian input error, we get
\begin{equation*}
    [\mathbf{K}_\eta'(0)]_{ij}=
  \begin{cases} \nabla^2c(s_i,s_j), & i\ne j,\\ 0, & i=j,\end{cases} \qquad \text{and} \qquad \left[\mathbf{k}^{*\prime}(0)\right]_i =
\frac{1}{2}\nabla^2c(s^*,s_i).
\end{equation*}
The diagonal derivatives are zero because the marginal observation variance, $[\mathbf{K}_\eta(\sigma_u^2)]_{ii}=(1+\eta)\tau^2$, does not depend on $\sigma_u^2$. The following theorem combines these covariance changes to determine the local change in prediction MSE under iid Gaussian input errors.

\begin{theorem}[Exact local criterion]
\label{thm:exact}
Suppose assumptions
(\ref{ass:smoothness_general})--(\ref{ass:pd_general}) hold.
Let
\begin{equation*}
\boldsymbol\alpha
=
\mathbf K_\eta(0)^{-1}\mathbf k^*(0)
\end{equation*}
denote the kriging weights at zero input-error variance, and define
\begin{equation*}
\Lambda
=
\sum_{i=1}^n
\alpha_i\nabla^2c(s^*,s_i)
-
\sum_{i\ne j}
\alpha_i\alpha_j\nabla^2c(s_i,s_j).
\end{equation*}
Then, as $\sigma_u^2\downarrow0$,
\begin{equation*}
M_{n,\eta}(\sigma_u^2)
=
M_{n,\eta}(0)
-
\Lambda\sigma_u^2
+
o(\sigma_u^2).
\end{equation*}
Equivalently,
\begin{equation*}
M_{n,\eta}'(0)=-\Lambda.
\end{equation*}
Consequently,
\begin{enumerate}
\item if $\Lambda>0$, there exists $\delta>0$ such that $M_{n,\eta}(\sigma_u^2)<M_{n,\eta}(0)$ for every $0<\sigma_u^2<\delta$;

\item if $\Lambda<0$, there exists $\delta>0$ such that $M_{n,\eta}(\sigma_u^2)>M_{n,\eta}(0)$ for every $0<\sigma_u^2<\delta$.
\end{enumerate}
When $\Lambda=0$, this first-order criterion is inconclusive.
\end{theorem}


\subsection{Proof of Theorem \ref{thm:exact}}
\label{Pf:ThmExact}

Since $c$ is stationary and isotropic, we write $c(h)=c(h,0)$ for a lag $h\in\mathbb{R}^p$. Hence, $c(s,t)=c(s-t)$, $c(-h)=c(h)$ and $\nabla^2c(s,t)=\nabla^2c(s-t)$. Where $c$ is twice differentiable, $\mathbf{H}_c(h)$ denotes its Hessian matrix at $h$, so that $\operatorname{tr}\bigl(\mathbf{H}_c(h)\bigr)=\nabla^2c(h)$. Throughout, $\|\cdot\|_2$ denotes the spectral norm when applied to a matrix; and for vectors and matrices, $O(\sigma_u^2)$ and $o(\sigma_u^2)$ are used element-wise. We prove Theorem \ref{thm:exact} step-by-step using four lemmas. Additionally, we assume $\tau^2=1$ without loss of generality. As a result, $|c(h)|\le c(0)=1$ for every $h\in\mathbb{R}^p$.


We first study how a small Gaussian perturbation changes the covariance function using the following lemma. Since the perturbation has mean zero, its linear contribution disappears after taking expectation. Therefore, the leading change is proportional to $\sigma^2$ and depends on the Laplacian of the covariance function at the original lag. We later use $\sigma=\sigma_u$ for the target--observation covariance and $\sigma=\sqrt{2}\sigma_u$ for the covariance between two distinct training observations.

\begin{lemma}[Small-variance expansion]
\label{lem:expansion}
Let $h\in\mathbb{R}^p$, and suppose that $c$ is twice
continuously differentiable on the closed ball
\begin{equation*}
B
=
\{x\in\mathbb{R}^p:\|x-h\|_2\le\rho\}
\end{equation*}
for some $\rho>0$. Then, as $\sigma\downarrow0$,
\begin{equation*}
\mathbb{E}\bigl[c(h+\sigma Z)\bigr]
=
c(h)
+
\frac{\sigma^2}{2}\nabla^2c(h)
+
o(\sigma^2).
\end{equation*}
\end{lemma}

\begin{proof}
Since the Gaussian perturbation has unbounded support, we use Taylor expansion restricted by the event
\begin{equation*}
\mathcal E_\sigma
=
\{\sigma\|Z\|_2\le\rho\}.
\end{equation*}
On $\mathcal E_\sigma$, $h+\sigma Z\in B$. Moreover, for
every $t\in[0,1]$, $\|h+t\sigma Z-h\|_2 = t\sigma\|Z\|_2 \le\rho$. Hence, the entire segment from $h$ to $h+\sigma Z$ lies in $B$. For a fixed realization of $Z$, define
\begin{equation*}
g(t)=c(h+t\sigma Z),
\qquad 0\le t\le1.
\end{equation*}
On $\mathcal E_\sigma$, Taylor's theorem with integral remainder therefore gives
\begin{equation*}
    g(1) = g(0) + g'(0) + \int_0^1 \int_0^t g''(r) dr dt.
\end{equation*}

Since $g'(t) =\sigma\nabla c(h+t\sigma Z)^\top Z$ and $g''(t) = \sigma^2 Z^\top\mathbf H_c(h+t\sigma Z)Z,$ using Fubini's theorem, we get

\begin{equation*}
\begin{aligned}
c(h+\sigma Z)
&=
c(h)+\sigma\nabla c(h)^\top Z +
\sigma^2\int_0^1(1-t)
Z^\top\mathbf H_c(h+t\sigma Z)Z\,dt.
\end{aligned}
\end{equation*}
Adding and subtracting $\mathbf H_c(h)$ inside the
integral, and using
$\int_0^1(1-t)\,dt=1/2$, we obtain
\begin{equation}
\label{eq:taylor_small_variance}
c(h+\sigma Z)
=
c(h)
+
\sigma\nabla c(h)^\top Z
+
\frac{\sigma^2}{2}
Z^\top\mathbf H_c(h)Z
+
\sigma^2R_\sigma,
\end{equation}
where, on $\mathcal E_\sigma$, 
\begin{equation*}
R_\sigma
=
\int_0^1(1-t)
Z^\top
\left[
\mathbf H_c(h+t\sigma Z)-\mathbf H_c(h)
\right]
Z\,dt.
\end{equation*}
For convenience, define $R_\sigma=0$ on $\mathcal E_\sigma^c$. Multiplying both sides of \eqref{eq:taylor_small_variance} by $\mathbf 1_{\mathcal E_\sigma}$  and then taking expectation, we get
\begin{align}
\mathbb E\left[
c(h+\sigma Z)\mathbf 1_{\mathcal E_\sigma}
\right]
={}&
c(h)\mathbb P(\mathcal E_\sigma)
+
\sigma\nabla c(h)^\top
\mathbb E\left[
Z\mathbf 1_{\mathcal E_\sigma}
\right]+
\frac{\sigma^2}{2}
\mathbb E\left[
Z^\top\mathbf H_c(h)Z
\mathbf 1_{\mathcal E_\sigma}
\right]
+
\sigma^2
\mathbb E\left[
R_\sigma\mathbf 1_{\mathcal E_\sigma}
\right].
\label{eq:restricted_taylor_expectation}
\end{align}

Using Markov's inequality,
\begin{align*}
\mathbb P(\mathcal E_\sigma^c)
&=
\mathbb P(\|Z\|_2>\rho/\sigma)
\nonumber\\
&\le
\frac{\sigma^4\mathbb E\|Z\|_2^4}{\rho^4}
=
O(\sigma^4).
\label{eq:tail_small_variance}
\end{align*}

Then the first component of the right hand side of \eqref{eq:restricted_taylor_expectation} becomes
\begin{align*}
c(h)\mathbb P(\mathcal E_\sigma)
&=
c(h)\left\{1-\mathbb P(\mathcal E_\sigma^c)\right\}\\
&=
c(h)+O(\sigma^4)\\
&=
c(h)+o(\sigma^2).
\end{align*}

Now, $Z\mathbf1_{\mathcal E_\sigma}$ is symmetric around zero since the event $\mathcal E_\sigma$ depends only on $\|Z\|_2$ and the distribution of $Z$ is symmetric. Hence, $\mathbb E\left[
Z\mathbf1_{\mathcal E_\sigma}\right] = \mathbf0$. As a result, the second term of the right hand side of \eqref{eq:restricted_taylor_expectation} becomes
\begin{equation*}
\mathbb E\left[ \nabla c(h)^\top Z \mathbf1_{\mathcal E_\sigma} \right]
=
0.
\end{equation*}

For the third component, we write
\begin{align*}
\mathbb E\left[
Z^\top\mathbf H_c(h)Z
\mathbf 1_{\mathcal E_\sigma}
\right]
={}&
\mathbb E\left[
Z^\top\mathbf H_c(h)Z
\right] -
\mathbb E\left[
Z^\top\mathbf H_c(h)Z
\mathbf 1_{\mathcal E_\sigma^c}
\right].
\end{align*}
The first expectation satisfies
\begin{align*}
\mathbb E\left[
Z^\top\mathbf H_c(h)Z
\right]
&=
\operatorname{tr}\left(
\mathbf H_c(h)\mathbb E[ZZ^\top]
\right)\\
&=
\operatorname{tr}\left(\mathbf H_c(h)\right)\\
&=
\nabla^2c(h).
\end{align*}
For the second expectation, using Jensen's and Cauchy--Schwarz inequality, we get
\begin{align*}
\left|
\mathbb E\left[
Z^\top\mathbf H_c(h)Z
\mathbf 1_{\mathcal E_\sigma^c}
\right]
\right|
&\le 
\mathbb E\left[ \left|
Z^\top\mathbf H_c(h)Z
\mathbf 1_{\mathcal E_\sigma^c}
\right| \right] \\
& \leq
\|\mathbf H_c(h)\|_F
\mathbb E\left[
\|Z\|_2^2\mathbf 1_{\mathcal E_\sigma^c}
\right]\\
&\le
\|\mathbf H_c(h)\|_F
\left(\mathbb E\|Z\|_2^4\right)^{1/2}
\mathbb P(\mathcal E_\sigma^c)^{1/2}\\
&=
O(\sigma^2).
\end{align*}
Hence,
\begin{equation*}
\frac{\sigma^2}{2}
\mathbb E\left[
Z^\top\mathbf H_c(h)Z
\mathbf 1_{\mathcal E_\sigma}
\right]
=
\frac{\sigma^2}{2}\nabla^2c(h)
+
O(\sigma^4) = \frac{\sigma^2}{2}\nabla^2c(h)
+
o(\sigma^2).
\end{equation*}

We next show that the expected contribution of $R_\sigma$ is negligible. Fix any $z\in\mathbb R^p$. If $z\ne\mathbf0$, then $\mathcal E_\sigma$ holds whenever
$\sigma\le\rho/\|z\|_2$; if $z=\mathbf0$, it holds for every $\sigma>0$. Thus, $\mathcal E_\sigma$ eventually holds for every fixed $z$ as $\sigma\downarrow0$.

Moreover,
\begin{equation*}
\sup_{t\in[0,1]}
\|h+t\sigma z-h\|_2
\le
\sigma\|z\|_2
\longrightarrow0.
\end{equation*}
By continuity of $\mathbf H_c$ at $h$,
\begin{equation*}
\sup_{t\in[0,1]}
\left\|
\mathbf H_c(h+t\sigma z)-\mathbf H_c(h)
\right\|_F
\longrightarrow0.
\end{equation*}
Therefore,
\begin{align*}
|R_\sigma|
&\le
\|z\|_2^2
\int_0^1(1-t)
\left\|
\mathbf H_c(h+t\sigma z)-\mathbf H_c(h)
\right\|_F\,dt\\
&\le
\frac12\|z\|_2^2
\sup_{t\in[0,1]}
\left\|
\mathbf H_c(h+t\sigma z)-\mathbf H_c(h)
\right\|_F
\longrightarrow0.
\end{align*}
It follows that
\begin{equation*}
R_\sigma\mathbf1_{\mathcal E_\sigma}
\longrightarrow0
\qquad\text{almost surely}.
\end{equation*}

Because $B$ is compact and $\mathbf H_c$ is continuous on $B$,
\begin{equation*}
H_{\max}
=
\sup_{x\in B}\|\mathbf H_c(x)\|_F
<\infty.
\end{equation*}
On $\mathcal E_\sigma$, both $h+t\sigma Z$ and $h$ belong to $B$. Hence, by the triangle inequality,
\begin{align*}
|R_\sigma|\mathbf1_{\mathcal E_\sigma}
&\le
\|Z\|_2^2
\int_0^1(1-t)
\left\{
\|\mathbf H_c(h+t\sigma Z)\|_F
+
\|\mathbf H_c(h)\|_F
\right\}dt\\
&\le
2H_{\max}\|Z\|_2^2
\int_0^1(1-t)\,dt\\
&=
H_{\max}\|Z\|_2^2.
\end{align*}
The dominating random variable is integrable because $\mathbb E\|Z\|_2^2=p < \infty$. Therefore, the almost sure variant of the dominated convergence theorem \citep{resnick1999probability} gives
\begin{equation*}
\label{eq:remainder_small_variance}
\mathbb E\left[
|R_\sigma|\mathbf1_{\mathcal E_\sigma}
\right]
\longrightarrow0.
\end{equation*}
Consequently,
\begin{align*}
\left|
\sigma^2
\mathbb E\left[
R_\sigma\mathbf1_{\mathcal E_\sigma}
\right]
\right|
&\le
\sigma^2
\mathbb E\left[
|R_\sigma|\mathbf1_{\mathcal E_\sigma}
\right]\\
&=
o(\sigma^2).
\end{align*}

Combining the four components in the right hand side of  \eqref{eq:restricted_taylor_expectation}, we obtain
\begin{equation}
\label{eq:restricted_small_variance}
\mathbb E\left[
c(h+\sigma Z)\mathbf 1_{\mathcal E_\sigma}
\right]
=
c(h)
+
\frac{\sigma^2}{2}\nabla^2c(h)
+
o(\sigma^2).
\end{equation}

It remains to consider the contribution from $\mathcal E_\sigma^c$. Since $c$ is a correlation function under the normalization $c(0)=1$, we have $|c(x)|\le1$. Therefore,
\begin{align*}
\left|
\mathbb E\left[
c(h+\sigma Z)\mathbf 1_{\mathcal E_\sigma^c}
\right]
\right|
&\le
\mathbb P(\mathcal E_\sigma^c)\\
&=
O(\sigma^4)\\
&=
o(\sigma^2).
\end{align*}

Finally,
\begin{align*}
\mathbb E[c(h+\sigma Z)]
=& 
\mathbb E\left[
c(h+\sigma Z)\mathbf 1_{\mathcal E_\sigma}
\right]+ \mathbb E\left[
c(h+\sigma Z)\mathbf 1_{\mathcal E_\sigma^c}
\right]\ \\
&= c(h)
+
\frac{\sigma^2}{2}\nabla^2c(h)
+
o(\sigma^2),
.
\end{align*}
which proves the result. In general, for any fixed inputs $s$ and $t$ such that $c()$ is twice continuously differentiable in a neighborhood of $h=s-t$, we get $c_{\sigma^2}(s,t)
= c(s,t) + \frac{\sigma^2}{2}\nabla^2c(s, t) + o(\sigma^2)$, as $\sigma^2 \downarrow 0$.

\end{proof}

\begin{lemma}[Expansion of the KALE covariances]
\label{lem:moments}
Define the vector $\mathbf{a}$ and the symmetric matrix $\mathbf{L}$ with elements
\begin{equation*}
a_i
=
\frac{1}{2}\nabla^2c(s^*-s_i),
\qquad
L_{ij}
=
\begin{cases}
\nabla^2c(s_i-s_j), & i\neq j, \\
0, & i=j.
\end{cases}
\end{equation*}
Under assumption \ref{ass:smoothness_general}, as $\sigma_u^2\downarrow0$,
\begin{align*}
\mathbf{k}^{*}(\sigma_u^2)
&=
\mathbf{k}^{*}(0) + \sigma_u^2\mathbf{a} + o(\sigma_u^2),\\
\mathbf{K}_{\eta}(\sigma_u^2)
&=
\mathbf{K}_{\eta}(0) + \sigma_u^2\mathbf{L} + o(\sigma_u^2).
\end{align*}
In particular, $\mathbf{k}^{*\prime}(0)=\mathbf{a}$ and $\mathbf{K}_{\eta}^{\prime}(0)=\mathbf{L}$.
\end{lemma}

\begin{proof}
For $i\neq j$, $u_i-u_j\sim N_p(\mathbf{0},2\sigma_u^2\mathbf{I}_p)$. Hence,
\begin{equation*}
[\mathbf{K}_{\eta}(\sigma_u^2)]_{ij}
=
\mathbb{E}\bigl[c(s_i-s_j+\sqrt{2}\,\sigma_uZ)\bigr].
\end{equation*}
Similarly, $-u_i\sim N_p(\mathbf{0},\sigma_u^2\mathbf{I}_p)$ gives
\begin{equation*}
[\mathbf{k}^{*}(\sigma_u^2)]_i
=
\mathbb{E}\bigl[c(s^*-s_i+\sigma_uZ)\bigr].
\end{equation*}
By assumption \ref{ass:smoothness_general}, Lemma \ref{lem:expansion} applies at every observation difference $s_i-s_j$, $i\neq j$, with $\sigma=\sqrt{2}\,\sigma_u$, and at every observation to target difference $s^*-s_i$ with $\sigma=\sigma_u$. Since $(\sqrt{2}\,\sigma_u)^2/2=\sigma_u^2$, this gives the off-diagonal elements of $\mathbf{L}$ and the elements of $\mathbf{a}$. The diagonal elements are $[\mathbf{K}_{\eta}(\sigma_u^2)]_{ii}=(1+\eta)$ for all $i \in [n]$, which do not depend on $\sigma_u^2$; hence, $L_{ii}=0$.
\end{proof}

\begin{lemma}[Covariance matrix remains invertible]
\label{lem:invertible}
Let $\lambda_0=\lambda_{\min}\bigl(\mathbf{K}_{\eta}(0)\bigr)$. Under assumptions \ref{ass:smoothness_general}--\ref{ass:pd_general}, there exists $\sigma_0>0$ such that, for every $0\le\sigma_u^2\le\sigma_0^2$,
\begin{equation*}
\mathbf{K}_{\eta}(\sigma_u^2)\succ0,
\qquad
\|\mathbf{K}_{\eta}(\sigma_u^2)^{-1}\|_2
\le
\frac{2}{\lambda_0},
\qquad
\|\mathbf{K}_{\eta}(\sigma_u^2)\|_2
\le
n(1+\eta).
\end{equation*}
\end{lemma}

\begin{proof}
By assumption \ref{ass:pd_general}, $\lambda_0>0$. By Lemma \ref{lem:moments},
\begin{equation*}
\|\mathbf{K}_{\eta}(\sigma_u^2)-\mathbf{K}_{\eta}(0)\|_F
\longrightarrow0
\qquad\text{as }\sigma_u^2\downarrow0.
\end{equation*}
Therefore, there exists $\sigma_0>0$ such that $\|\mathbf{K}_{\eta}(\sigma_u^2)-\mathbf{K}_{\eta}(0)\|_F
\le
\frac{\lambda_0}{2}$ for every $0\le\sigma_u^2\le\sigma_0^2$. Weyl's inequality \citep[Theorem~4.3.1]{horn2013matrix} gives
\begin{align*}
\lambda_{\min}\bigl(\mathbf{K}_{\eta}(\sigma_u^2)\bigr)
&\ge
\lambda_{\min}\bigl(\mathbf{K}_{\eta}(0)\bigr)
-
\|\mathbf{K}_{\eta}(\sigma_u^2)-\mathbf{K}_{\eta}(0)\|_F\\
&\ge
\frac{\lambda_0}{2}
>0.
\end{align*}
Hence, $\mathbf{K}_{\eta}(\sigma_u^2)\succ0$ and $\|\mathbf{K}_{\eta}(\sigma_u^2)^{-1}\|_2=1/\lambda_{\min}\bigl(\mathbf{K}_{\eta}(\sigma_u^2)\bigr)\le2/\lambda_0$. Finally, every element of $\mathbf{K}_{\eta}(\sigma_u^2)$ is at most $1+\eta$ in absolute value. Hence, $\|\mathbf{K}_{\eta}(\sigma_u^2)\|_2\le\|\mathbf{K}_{\eta}(\sigma_u^2)\|_F\le n(1+\eta)$.
\end{proof}

\begin{lemma}[Exact decomposition of the MSE]
\label{lem:decomposition}
For $0\le\sigma_u^2\le\sigma_0^2$, define the KALE kriging-weight vector by
\begin{equation}
\boldsymbol{\alpha}_{\eta}(\sigma_u^2)
=
\mathbf{K}_{\eta}(\sigma_u^2)^{-1}
\mathbf{k}^{*}(\sigma_u^2),
\end{equation}
and write $\boldsymbol{\alpha}=\boldsymbol{\alpha}_{\eta}(0)$. Then
\begin{align}
M_{n,\eta}(0)-M_{n,\eta}(\sigma_u^2)
&=
2\boldsymbol{\alpha}^\top
\bigl\{\mathbf{k}^{*}(\sigma_u^2)-\mathbf{k}^{*}(0)\bigr\}
-
\boldsymbol{\alpha}^\top
\bigl\{\mathbf{K}_{\eta}(\sigma_u^2)-\mathbf{K}_{\eta}(0)\bigr\}
\boldsymbol{\alpha}
\nonumber\\
&\quad+
\bigl\{\boldsymbol{\alpha}_{\eta}(\sigma_u^2)-\boldsymbol{\alpha}\bigr\}^\top
\mathbf{K}_{\eta}(\sigma_u^2)
\bigl\{\boldsymbol{\alpha}_{\eta}(\sigma_u^2)-\boldsymbol{\alpha}\bigr\}.
\label{eq:mse-decomposition}
\end{align}
\end{lemma}

\begin{proof}
Define the quadratic function
\begin{equation*}
Q(\mathbf{w})
=
2\mathbf{w}^\top\mathbf{k}^{*}(\sigma_u^2)
-
\mathbf{w}^\top\mathbf{K}_{\eta}(\sigma_u^2)\mathbf{w},
\qquad
\mathbf{w}\in\mathbb{R}^n.
\end{equation*}
By Lemma \ref{lem:invertible}, $\mathbf{K}_{\eta}(\sigma_u^2)\succ0$. Hence, $Q$ is strictly concave and is maximized at $\boldsymbol{\alpha}_{\eta}(\sigma_u^2)$, with maximum
\begin{equation*}
Q\bigl(\boldsymbol{\alpha}_{\eta}(\sigma_u^2)\bigr)
=
\mathbf{k}^{*}(\sigma_u^2)^\top
\mathbf{K}_{\eta}(\sigma_u^2)^{-1}
\mathbf{k}^{*}(\sigma_u^2)
=
1-M_{n,\eta}(\sigma_u^2).
\end{equation*}
Expanding $Q$ around its maximizer gives
\begin{equation*}
Q\bigl(\boldsymbol{\alpha}_{\eta}(\sigma_u^2)\bigr)
-
Q(\boldsymbol{\alpha})
=
\bigl\{\boldsymbol{\alpha}_{\eta}(\sigma_u^2)-\boldsymbol{\alpha}\bigr\}^\top
\mathbf{K}_{\eta}(\sigma_u^2)
\bigl\{\boldsymbol{\alpha}_{\eta}(\sigma_u^2)-\boldsymbol{\alpha}\bigr\}.
\end{equation*}
Since $\mathbf{K}_{\eta}(0)\boldsymbol{\alpha}=\mathbf{k}^{*}(0)$, we have $\boldsymbol{\alpha}^\top\mathbf{k}^{*}(0)=\boldsymbol{\alpha}^\top\mathbf{K}_{\eta}(0)\boldsymbol{\alpha}=1-M_{n,\eta}(0)$. Hence,
\begin{equation*}
Q(\boldsymbol{\alpha})
=
1-M_{n,\eta}(0)
+
2\boldsymbol{\alpha}^\top
\bigl\{\mathbf{k}^{*}(\sigma_u^2)-\mathbf{k}^{*}(0)\bigr\}
-
\boldsymbol{\alpha}^\top
\bigl\{\mathbf{K}_{\eta}(\sigma_u^2)-\mathbf{K}_{\eta}(0)\bigr\}
\boldsymbol{\alpha}.
\end{equation*}
Combining the last three displays gives \eqref{eq:mse-decomposition}.
\end{proof}

Now we shall finalize the proof of Theorem \ref{thm:exact} using Lemma \ref{lem:expansion}--\ref{lem:decomposition}
\begin{proof}[Proof of Theorem \ref{thm:exact}]
Let $0<\sigma_u^2\le\sigma_0^2$, with $\sigma_0$ from Lemma \ref{lem:invertible}. We consider the three terms in \eqref{eq:mse-decomposition} in turn.

By Lemma \ref{lem:moments}, the first two terms satisfy
\begin{align*}
2\boldsymbol{\alpha}^\top
\bigl\{\mathbf{k}^{*}(\sigma_u^2)-\mathbf{k}^{*}(0)\bigr\}
-
\boldsymbol{\alpha}^\top
\bigl\{\mathbf{K}_{\eta}(\sigma_u^2)-\mathbf{K}_{\eta}(0)\bigr\}
\boldsymbol{\alpha}
&=
\sigma_u^2
\bigl(
2\boldsymbol{\alpha}^\top\mathbf{a}
-
\boldsymbol{\alpha}^\top\mathbf{L}\boldsymbol{\alpha}
\bigr)
+
o(\sigma_u^2)\\
&=
\sigma_u^2\Lambda
+
o(\sigma_u^2),
\end{align*}
since
\begin{equation*}
2\boldsymbol{\alpha}^\top\mathbf{a}
-
\boldsymbol{\alpha}^\top\mathbf{L}\boldsymbol{\alpha}
=
\sum_{i=1}^n\alpha_i\nabla^2c(s^*-s_i)
-
\sum_{i\neq j}\alpha_i\alpha_j\nabla^2c(s_i-s_j)
=
\Lambda.
\end{equation*}

For the last term, $\mathbf{K}_{\eta}(0)\boldsymbol{\alpha}=\mathbf{k}^{*}(0)$ gives
\begin{align*}
\boldsymbol{\alpha}_{\eta}(\sigma_u^2)-\boldsymbol{\alpha}
&=
\mathbf{K}_{\eta}(\sigma_u^2)^{-1}
\bigl\{\mathbf{k}^{*}(\sigma_u^2)-\mathbf{K}_{\eta}(\sigma_u^2)\boldsymbol{\alpha}\bigr\}\\
&=
\mathbf{K}_{\eta}(\sigma_u^2)^{-1}
\Bigl[
\bigl\{\mathbf{k}^{*}(\sigma_u^2)-\mathbf{k}^{*}(0)\bigr\}
-
\bigl\{\mathbf{K}_{\eta}(\sigma_u^2)-\mathbf{K}_{\eta}(0)\bigr\}\boldsymbol{\alpha}
\Bigr].
\end{align*}
By Lemmas \ref{lem:moments} and \ref{lem:invertible}, $\|\boldsymbol{\alpha}_{\eta}(\sigma_u^2)-\boldsymbol{\alpha}\|_2=O(\sigma_u^2)$. Hence,
\begin{equation*}
0
\le
\bigl\{\boldsymbol{\alpha}_{\eta}(\sigma_u^2)-\boldsymbol{\alpha}\bigr\}^\top
\mathbf{K}_{\eta}(\sigma_u^2)
\bigl\{\boldsymbol{\alpha}_{\eta}(\sigma_u^2)-\boldsymbol{\alpha}\bigr\}
\le
n(1+\eta)
\|\boldsymbol{\alpha}_{\eta}(\sigma_u^2)-\boldsymbol{\alpha}\|_2^2
=
O(\sigma_u^4).
\end{equation*}

Therefore,
\begin{equation*}
M_{n,\eta}(\sigma_u^2)-M_{n,\eta}(0)
=
-\Lambda\sigma_u^2
+
o(\sigma_u^2)
\qquad\text{as }\sigma_u^2\downarrow0.
\end{equation*}
Hence, $M_{n,\eta}$ has a right derivative at $\sigma_u^2=0$, and $M_{n,\eta}^{\prime}(0)=-\Lambda$.

Suppose $\Lambda>0$. Choose $\delta\in(0,\sigma_0^2]$ such that the $o(\sigma_u^2)$ term is at most $\Lambda\sigma_u^2/2$ in absolute value for every $0<\sigma_u^2<\delta$. Then
\begin{equation*}
M_{n,\eta}(\sigma_u^2)-M_{n,\eta}(0)
\le
-\frac{\Lambda\sigma_u^2}{2}
<0
\qquad
\text{for every }0<\sigma_u^2<\delta.
\end{equation*}
If $\Lambda<0$, the same argument gives $M_{n,\eta}(\sigma_u^2)-M_{n,\eta}(0)\ge|\Lambda|\sigma_u^2/2>0$ for every $0<\sigma_u^2<\delta$.
\end{proof}

The quantity $\Lambda$ in theorem \ref{thm:exact} describes the balance between two first-order effects:
\begin{equation*}
\Lambda
=
\underbrace{
2\boldsymbol\alpha^\top
\mathbf k^{*\,\prime}(0)
}_{\textnormal{change in target--predictor covariance}}
-
\underbrace{
\boldsymbol\alpha^\top
\mathbf K_\eta'(0)\boldsymbol\alpha
}_{\textnormal{change in predictor variance}}.
\end{equation*}

With the weights $\boldsymbol\alpha$ held at their zero-input-error values, the first term measures twice the change in covariance between the prediction target and the weighted observations. The second term measures the change in the variance of those weighted observations. When $\Lambda>0$, the first term exceeds the second, giving $M_{n,\eta}'(0)<0$ and a local reduction in prediction MSE. When $\Lambda<0$, the comparison is reversed and prediction MSE locally increases. When $\Lambda=0$, the first-order expansion does not determine the local behavior.

Now, if the covariance function is non-zero, then we can further represent $\Lambda$ as a quadratic form.
\begin{lemma}[Quadratic-form representation]
\label{lem:lambda_quadratic}
Suppose (\ref{ass:smoothness_general})--(\ref{ass:pd_general}) hold and $c(s^*,s_i)\ne0$ for every $i\in[n]$.
Let $q_i=\nabla^2c(s^*,s_i)/c(s^*,s_i)$. Then
$\boldsymbol\alpha\ne\mathbf 0$ and
$\Lambda=\boldsymbol\alpha^\top \mathbf A_\eta\boldsymbol\alpha$,
where $\mathbf A_\eta$ is symmetric with entries
\begin{equation*}
[\mathbf A_\eta]_{ij}
=
\begin{cases}
(1+\eta)\tau^2q_i, & i=j,\\[4pt]
\displaystyle\frac{q_i+q_j}{2}\,c(s_i,s_j)
-\nabla^2c(s_i,s_j), & i\ne j.
\end{cases}
\end{equation*}
In particular, $\mathbf A_\eta\succ0$ implies $\Lambda>0$.
\end{lemma}

\begin{proof}
By assumption~\ref{ass:pd_general}, $\mathbf K_\eta(0)\succ0$.
Since $\mathbf K_\eta(0)\boldsymbol\alpha=\mathbf k^*(0)$,
\begin{equation*}
\|\boldsymbol\alpha\|_2^2
\ge
\frac{\|\mathbf k^*(0)\|_2^2}
     {\lambda_{\max}^2(\mathbf K_\eta(0))}
>0,
\end{equation*}
where the strict inequality follows from
$[\mathbf k^*(0)]_i=c(s^*,s_i)\ne0$. Thus
$\boldsymbol\alpha\ne\mathbf 0$.

Using the definition of $q_i$ and the symmetry of
$\mathbf K_\eta(0)$, the target term in $\Lambda$ satisfies
\begin{align*}
\sum_{i=1}^n\alpha_i\nabla^2c(s^*,s_i)
&=\sum_{i=1}^n\alpha_i q_i[\mathbf k^*(0)]_i\\
&=\sum_{i,j=1}^n\alpha_i\alpha_j q_i[\mathbf K_\eta(0)]_{ij}\\
&=\sum_{i,j=1}^n\alpha_i\alpha_j
  \frac{q_i+q_j}{2}[\mathbf K_\eta(0)]_{ij}.
\end{align*}
The diagonal entries of $\mathbf K_\eta(0)$ are
$(1+\eta)\tau^2$, and its off-diagonal entries are $c(s_i,s_j)$.
Subtracting the off-diagonal Laplacian term in $\Lambda$ gives
$\Lambda=\boldsymbol\alpha^\top\mathbf A_\eta\boldsymbol\alpha$
with the stated entries. The matrix $\mathbf A_\eta$ is symmetric
because the Laplacian of a stationary covariance is even in the lag.
Since $\boldsymbol\alpha\ne\mathbf 0$, positive definiteness of
$\mathbf A_\eta$ implies $\Lambda>0$.
\end{proof}

Under the same nonnegativity of the covariance, we can show that the local improvement of MSE does not hold for arbitrary large input error variance. To show this, we impose one additional assumption.
\begin{assumptions}[
label=\textnormal{(G\arabic*)}, ref=G\arabic*, start=3
]
\item \label{ass:cov_decay}
The covariance function decays to zero; meaning $c(h)\to0$ as $\|h\|_2\to\infty$.
\end{assumptions}

For a stationary GP with continuous covariance, assumption~(\ref{ass:cov_decay}) implies ergodicity \citep{ergodic_suff}; however, it is stronger and holds for the SE Mat\'ern (including exponential) and rational-quadratic covariance families.

\begin{lemma}[Prediction MSE for large input-error variance]
\label{lem:large_input_error}
Under the iid Gaussian input-error model, suppose (\ref{ass:smoothness_general})--(\ref{ass:cov_decay}) hold.
Then
\begin{equation*}
\lim_{\sigma_u^2\to\infty}M_{n,\eta}(\sigma_u^2)=\tau^2.
\end{equation*}
If $\mathbf k^*(0)\ne\mathbf 0$, then
$M_{n,\eta}(\sigma_u^2)>M_{n,\eta}(0)$ for all sufficiently
large $\sigma_u^2$. In particular, if (\ref{ass:smoothness_general}) also holds and $\Lambda>0$, the local improvement established in Theorem~\ref{thm:exact} cannot persist for arbitrarily large
input-error variances.

\end{lemma}

\begin{proof}
Let $Z\sim N_p(\mathbf 0,\mathbf I_p)$. Since
$\mathbb P(Z=\mathbf 0)=0$, for every fixed
$s,t\in\mathbb R^p$,
\begin{equation*}
\|s-t+\sigma  Z\|_2\longrightarrow\infty
\qquad\text{almost surely as }\sigma\to\infty.
\end{equation*}
Assumption~(\ref{ass:cov_decay}) therefore gives
\begin{equation*}
c(s-t+\sqrt \sigma Z)\longrightarrow0
\qquad\text{almost surely}.
\end{equation*}
Moreover, by the Cauchy--Schwarz inequality, the stationary
covariance satisfies $|c(h)|\le c(0)=\tau^2$ for every
$h\in\mathbb R^p$. Using dominated convergence theorem, we get
\begin{equation*}
c_{\sigma^2}(s,t)
:=
\mathbb E[c(s-t+\sigma Z)]
\longrightarrow0
\qquad\text{as }\sigma^2\to\infty.
\end{equation*}

Using this, we get $[\mathbf k^*(\sigma_u^2)]_i = c_{\sigma_u^2}(s_i,s^*)\longrightarrow0,$ and $[\mathbf K_\eta(\sigma_u^2)]_{ij}
 = c_{2 \sigma_u^2}(s_i,s_j)\longrightarrow0$ for $i \neq j$ when $\sigma_u^2 \longrightarrow \infty$. The diagonal entries remain equal to $(1+\eta)\tau^2$. Since $n$ is finite, we thus have
\begin{equation*}
\mathbf k^*(\sigma_u^2)\longrightarrow\mathbf 0,
\qquad
\mathbf K_\eta(\sigma_u^2)
\longrightarrow
(1+\eta)\tau^2\mathbf I_n\succ0.
\end{equation*}
Consequently, $\mathbf K_\eta(\sigma_u^2)$ is positive definite
for all sufficiently large $w$, and continuity of matrix
inversion gives
\begin{equation*}
\mathbf K_\eta(\sigma_u^2)^{-1}
\longrightarrow
\frac{1}{(1+\eta)\tau^2}\mathbf I_n.
\end{equation*}
It follows that
\begin{equation*}
M_{n,\eta}(\sigma_u^2)
=
\tau^2-\mathbf k^*(\sigma_u^2)^\top
\mathbf K_\eta(\sigma_u^2)^{-1}\mathbf k^*(\sigma_u^2)
\longrightarrow\tau^2.
\end{equation*}

If $\mathbf k^*(0)\ne\mathbf 0$, then
$\mathbf K_\eta(0)\succ0$ implies
\begin{equation*}
M_{n,\eta}(0)
=
\tau^2-\mathbf k^*(0)^\top
\mathbf K_\eta(0)^{-1}\mathbf k^*(0)
<
\tau^2.
\end{equation*}
Since the limiting MSE is strictly larger than the baseline,
there exists $w_0<\infty$ such that
\begin{equation*}
M_{n,\eta}(\sigma_u^2)>M_{n,\eta}(0)
\qquad\text{for every }\sigma_u^2>w_0.
\end{equation*}
Finally, $\Lambda>0$ implies $\mathbf k^*(0)\ne\mathbf 0$,
since $\Lambda$ is a quadratic form in
$\mathbf K_\eta(0)^{-1}\mathbf k^*(0)$. Thus, the local
improvement established in Theorem~\ref{thm:exact} cannot
persist for arbitrarily large input-error variances.
\end{proof}

\section{Covariance-specific conditions for local MSE improvement}

\subsection{SE-GE model}
Now we focus on the SE--GE model and derive distance-based conditions that ensures local improvement of MSE, ie $\Lambda>0$ in theorem \ref{thm:exact}. At first we consider the no additive observation error case. First, define the scaled squared target and pairwise distances by $D_i=\|s_i-s^*\|_2^2/\ell^2$ and $H_{ij}=\|s_i-s_j\|_2^2/\ell^2$, respectively. We then impose the following assumptions.

\begin{assumptions}

\item \label{ass:distinct}
The nominal training locations $s_1,\ldots,s_n$ are pairwise distinct.


\item \label{ass:distance-condition}
\begin{equation}
\tag{C}\label{cond:C}
\begin{aligned}
2D_i-p &> 0,
&& i\in[n],\\
(2D_i-p)(2D_j-p)
&>
\prod_{l\in\{i,j\}}
\sum_{k\neq l}
e^{-H_{lk}}
\left|D_l+D_k-2H_{lk}\right|,
&& 1\le i<j\le n.
\end{aligned}
\end{equation}

\end{assumptions}

\begin{proposition}
\label{thm: kale_no_nugget}
Let the KALE prediction MSE under the SE--GE model without additive observation noise be
$M_n(\sigma_u^2)=\tau^2-
\mathbf{k}^{*}(\sigma_u^2)^\top
\mathbf{K}(\sigma_u^2)^{-1}
\mathbf{k}^{*}(\sigma_u^2)$.
Under assumptions~\ref{ass:distinct}--\ref{ass:distance-condition},
there exists $\delta>0$ such that
\begin{equation*}
M_n(\sigma_u^2)<M_n(0)
\qquad
\text{for every }0<\sigma_u^2<\delta.
\end{equation*}
\end{proposition}

For the SE covariance, Assumption~\ref{ass:distinct} ensures $\mathbf K(0)\succ0$, as required by Assumption~\ref{ass:pd_general} of Theorem~\ref{thm:exact}. Condition~(\ref{cond:C}) is a sufficient condition that balances the target-distance contributions against the interactions among the training observations. The first inequality in \eqref{cond:C}, $D_i>p/2$, requires each nominal training location to be far enough from the target that a small increase in Gaussian input-error variance from zero increases their covariance. With multiple observations, Gaussian input error also changes the covariances
among observations. The second inequality controls how these changes combine in the MSE derivative by requiring, for each pair, the product of its positive diagonal contributions to exceed the product of the corresponding absolute off-diagonal row sums. When $n=1$, condition~(\ref{cond:C}) reduces to $D_1>p/2$, which is exactly the condition developed by \citet{cervone2015loc_error}.


\label{Pf:Thm1}

\begin{proof}

From lemma \ref{lem:lambda_quadratic}, we have $\boldsymbol\alpha\ne\mathbf0$. To establish $\Lambda>0$, we only need to show that
$\mathbf A_0 \succ 0$.

The symmetric matrix $\mathbf A_0$ in Lemma~\ref{lem:lambda_quadratic}
has diagonal entries $2D_i-p$ and absolute off-diagonal row sums
\begin{equation*}
R_i(0)=\sum_{j\ne i}e^{-H_{ij}}\left|D_i+D_j-2H_{ij}\right|.
\end{equation*}
For $n=1$, $2D_1-p>0$ ensures $\mathbf A_0\succ0$. Now suppose $n\ge2$ and $\mathbf A_0$ has an eigenvalue $\lambda\le0$. For $i\ne j$, Brauer's theorem \citep{horn2013matrix} provides
\begin{equation*}
R_i(0)R_j(0)
\ge |\lambda-a_i|\,|\lambda-a_j|
\ge a_i a_j
>R_i(0)R_j(0),
\end{equation*}
which is a contradiction. Hence, $\mathbf A_0\succ0$. 
\end{proof}


We next consider the SE--GE model with independent additive
observation noise and a fixed noise-to-signal variance ratio
$\eta>0$. Here we impose the following sufficient condition.

\begin{assumptions}[
label=\textnormal{(B\arabic*)},
ref=B\arabic*
]

\item \label{ass:distance-condition-nugget}
\begin{equation}
\tag{D}\label{cond:D}
\begin{aligned}
2D_i-p &> 0,
&& i\in[n],\\
(1+\eta)^2(2D_i-p)(2D_j-p)
&>
\prod_{l\in\{i,j\}}
\left(
\sum_{k\neq l}
e^{-H_{lk}}
\left|D_l+D_k-2H_{lk}\right|
\right),
&& 1\le i<j\le n.
\end{aligned}
\end{equation}

\end{assumptions}

A positive observation-noise variance ensures that the observation covariance matrix is invertible, even when some nominal training inputs are repeated. As shown by \citet{cervone2015loc_error}, the smoothed covariance $c_{\sigma^2}$ in equation~\eqref{eq: genenral_expectation} is a valid covariance function, resulting $\mathbf K(\sigma_u^2) \succeq 0$. Since $\eta>0$, $\mathbf K_\eta(\sigma_u^2) =\mathbf K(\sigma_u^2)+\eta\tau^2\mathbf I_n \succ 0$. Hence, unlike the setting without additive observation noise, the nominal training inputs are not required to
be distinct. 

Condition~(\ref{cond:D}) differs from condition~(\ref{cond:C}) only through the factor $(1+\eta)^2$ on the left-hand side of the second inequality. This factor represents the additional diagonal contribution from the observation noise and relaxes the sufficient condition for local improvement.

\begin{proposition}
\label{thm: kale_nugget}
Suppose $\eta>0$. Under
assumption~\ref{ass:distance-condition-nugget}, there exists
$\delta>0$ such that
\begin{equation*}
M_{n,\eta}(\sigma_u^2)<M_{n,\eta}(0)
\qquad
\text{for every }0<\sigma_u^2<\delta.
\end{equation*}
\end{proposition}

Since $(1+\eta)^2>1$, every input configuration satisfying condition~(\ref{cond:C}) also satisfies
condition~(\ref{cond:D}), while condition~(\ref{cond:D}) may hold
when condition~(\ref{cond:C}) fails. Thus, additive observation
noise expands the class of input configurations covered by the
sufficient condition. Nevertheless, both conditions are only
sufficient. Failure to satisfy either condition does not rule out
local improvement in the KALE prediction MSE.

\begin{proof}
The quadratic-form calculation from Proposition~\ref{thm: kale_no_nugget} applies with $\mathbf K_\eta$ in place of $\mathbf K$. The off-diagonal entries of the resulting matrix $\mathbf A_{\eta}$ are unchanged, while its diagonal entries are $(1+\eta)(2D_i-p)$, instead of $(2D_i-p)$. The condition~\eqref{cond:D} gives $\mathbf A_{\eta}\succ0$ by the same Brauer argument for $n\ge2$ and directly for $n=1$. 
\end{proof}



\subsection{Exp-GE model}

The distance conditions for the SE--GE model arise from how the
curvature of the SE covariance changes with the lag. We now consider
the one-dimensional Exp--GE model, for which the covariance has a
simpler curvature away from zero:
\begin{equation*}
\frac{d^2}{dr^2}c(r)
=
\frac{1}{\ell^2}c(r)>0,
\qquad r\ne0.
\end{equation*}
Using this relation into Theorem~\ref{thm:exact}, the target and off-diagonal training contributions cancel, leaving a strictly positive diagonal
contribution. Consequently, local MSE improvement does not require a lower bound on the target distances or pairwise training separations.

The exponential covariance is not differentiable at $r=0$. Therefore, the preceding argument applies only when all target and off-diagonal training lags are nonzero. We impose the following assumptions.

\begin{assumptions}[
label=\textnormal{(E\arabic*)},
ref=E\arabic*
]

\item
\label{ass:exp_diff_train}
The nominal training locations $s_1,\ldots,s_n$ are pairwise distinct.

\item
\label{ass:exp_diff_test}
The prediction target differs from every nominal training location; meaning $s^*\notin\{s_1,\ldots,s_n\}$.
\end{assumptions}

Unlike the SE covariance, the training locations must be
distinct regardless of whether additive observation noise is
present. The target must also differ from every training location.
These restrictions are required because the exponential covariance
function is not differentiable at zero lag.

\begin{proposition}[Local improvement under one-dimensional exponential covariance]
\label{thm:exponential_local_improvement}
Under assumptions~\ref{ass:exp_diff_train}--\ref{ass:exp_diff_test},
for any fixed $\eta\ge0$, there exists $\delta>0$ such that
\begin{equation*}
M_{n,\eta}(\sigma_u^2)
<
M_{n,\eta}(0)
\qquad
\text{for every }0<\sigma_u^2<\delta.
\end{equation*}
\end{proposition}

This result contrasts with the SE--GE case, where the sufficient
conditions depend on the distances among the training locations
and from the training locations to the target. For the
one-dimensional exponential covariance, every finite collection of distinct training locations and every unobserved target satisfies the local improvement criterion. A similar result can be obtained for a separable exponential covariance function in higher dimensions under coordinatewise distinctness conditions. However, this covariance function is not isotropic with respect to Euclidean distance when $p\ge2$. We therefore restrict attention to the one-dimensional exponential covariance model.

\begin{proof}
For the one-dimensional exponential covariance, $c(r)=\ell^2c''(r)$ for $h\ne0$. Since $s^*\ne s_i$, this gives $q_i=\nabla^2c(s^*,s_i)/c(s^*,s_i)=1/\ell^2$ for every $i$. Similarly, $s_i\ne s_j$ gives $\nabla^2c(s_i,s_j)=c(s_i,s_j)/\ell^2$ for $i\ne j$. Then the elements of $\mathbf A_\eta$ in Lemma~\ref{lem:lambda_quadratic} become
\begin{align*}
[\mathbf A_\eta]_{ii}
&=(1+\eta)\tau^2q_i
=\frac{(1+\eta)\tau^2}{\ell^2},\\
[\mathbf A_\eta]_{ij}
&=\frac{q_i+q_j}{2}\,c(s_i,s_j)-\nabla^2c(s_i,s_j)\\
&=\frac{c(s_i,s_j)}{\ell^2}-\frac{c(s_i,s_j)}{\ell^2}
=0,\qquad i\ne j.
\end{align*}
Hence we have $\mathbf A_\eta=\frac{(1+\eta)\tau^2}{\ell^2}\mathbf I_n\succ0$.
\end{proof}

\section{Example: Sparsely sampled time series}

Sparsely and irregularly sampled time series often arise in biological applications, where GPs are used to analyze temporal variation from a small number of observations. For example, \citet{topa2015short} showed that a very short estimated lengthscale for a GP with SE covariance could result in overfitting and recommended imposing lengthscale constraints based on the sampling intervals. \citet{topa2015sequencing} adopted a corresponding lower bound when analyzing sequencing time series, while \citet{somani2020personalised} used an informative lengthscale prior to discourage rapid fluctuations in longitudinal gene-expression models.

For time series data, proposition~\ref{thm:exponential_local_improvement} showed the existence of local improvement for every unobserved target and every finite collection of distinct training locations for Exp--GE model.
Focusing on the SE--GP model, in this section, we derive simplified ``task-oriented'' sufficient conditions for local improvement in prediction MSE based on conditions~(\ref{cond:C}) and~(\ref{cond:D}) for sparsely sampled time-series data. We first show that if the sampled times are sparse and the forecasting timepoint is appropriately far enough from the last observation time, local improvement for MSE holds for SE-GE model.

\begin{corollary}[Sparse time-series forecasting using SE-GE model]
\label{corr: one_dim_extra_no_nugget}
Let $\ell>0$ and let $s_i=(i-1)k$, $i=1,\ldots,n$, where $n\ge2$ and $k>0$. Define $h=k/\ell$, and suppose $h\ge0.9$.
We define the following quantities
\begin{equation*}
\begin{aligned}
D_0(h,n)
&=
\frac{2e^{-h^2}\left[1-e^{-3(n-1)h^2}\right]}
     {1-e^{-3h^2}},\\
D_2(h,n)
&=
\frac{
2h^2e^{-h^2}
\left[1-\left(4e^{-3h^2}\right)^{n-1}\right]
}
{1-4e^{-3h^2}}.
\end{aligned}
\end{equation*}
Suppressing the arguments $(h,n)$ of $D_0$ and $D_2$, define
\begin{equation*}
\begin{aligned}
d(h,n)=\frac{1}{2h}\Bigg[
&\Bigg(
h^2+\frac{2(1+D_0D_2)}{1-D_0^2}
+
\frac{
2\sqrt{(D_0+D_2)^2+
2h^2(1-D_0^2)(1+D_0D_2)}
}
{1-D_0^2}
\Bigg)^{1/2}-h
\Bigg].
\end{aligned}
\end{equation*}
Then every finite prediction target satisfying
\begin{equation}
s^*\ge s_n+d(h,n)k
\end{equation}
satisfies condition~(\ref{cond:C}). In particular, for a target $s^*=s_n+qk$ with a positive integer $q$ satisfying $q\ge\lceil d(h,n)\rceil$, there exists $\delta>0$ such that the KALE prediction MSE, $M_n(\sigma_u^2)$, is strictly decreasing in $\sigma_u^2$ on $(0,\delta)$.
\end{corollary}

\begin{proof}
    See Section \ref{Pf:cor1}.
\end{proof}

\begin{remark}
The condition $h\ge0.9$ characterizes sparsity relative to the covariance lengthscale; it does not imply a small sample size. Such configurations can arise when the sampling interval equals or exceeds the temporal dependence scale. For example, the gene-expression dataset GSE10562 \citep{dellagatta2008targets} contains observations at $20$-minute intervals, while \citet{kalaitzis2011simple} used a GP model with SE covariance and an initialization $\ell^2=40$. Under this initialization, $h=20/\sqrt{40}>0.9$.
\end{remark}

Several selected values of $d(h,n)$ are provided in Table~\ref{tab:values_of_c}. The bound is increasing in $n$ and decreasing in $h$. The limiting case $n\to\infty$ gives a universal upper bound on the sufficient extrapolation distance, expressed in grid intervals. The values stabilize rapidly as $n$ increases, particularly for larger $h$, suggesting that the grid spacing relative to the lengthscale has a greater influence on the bound than the sample size in the settings considered. For example, when $k/\ell\ge1.15$, the bound is smaller than one for every finite $n$. Consequently, every forecast target one or more grid steps beyond the final observation satisfies the sufficient condition. Under the stated model assumptions, the KALE prediction MSE at each such target is therefore lower than its zero-input-error value for all sufficiently small positive input-error variances.

\begin{table}
\centering
\caption{Values of $d(h,n)$ for selected scaled grid spacings $h=k/\ell$ and training sample sizes $n$, rounded to two decimal places. The column $n=\infty$ gives the limiting bound, valid for every finite $n$.}
\label{tab:values_of_c}
\begin{tabular}{@{}ccccccc@{}}
\toprule
& \multicolumn{6}{c}{Training sample size $n$} \\
\cmidrule(lr){2-7}
$h$ & $2$ & $5$ & $10$ & $25$ & $100$ & $\infty$ \\
\midrule
0.90 & 2.72 & 6.82 & 6.86 & 6.86 & 6.86 & 6.86 \\
1.00 & 1.50 & 1.73 & 1.73 & 1.73 & 1.73 & 1.73 \\
1.10 & 1.06 & 1.11 & 1.11 & 1.11 & 1.11 & 1.11 \\
1.15 & 0.93 & 0.96 & 0.96 & 0.96 & 0.96 & 0.96 \\
1.50 & 0.52 & 0.52 & 0.52 & 0.52 & 0.52 & 0.52 \\
1.75 & 0.41 & 0.41 & 0.41 & 0.41 & 0.41 & 0.41 \\
2.00 & 0.36 & 0.36 & 0.36 & 0.36 & 0.36 & 0.36 \\
3.00 & 0.24 & 0.24 & 0.24 & 0.24 & 0.24 & 0.24 \\
\bottomrule
\end{tabular}
\end{table}

Now instead of forecasting, suppose that our goal is to impute a missing value. The following corollary provides a sufficient condition for local improvement in the corresponding prediction MSE.

\begin{corollary}[Sparse time-series missing value imputation]
\label{corr: imputing_missing}
Consider the grid and notation of Corollary~\ref{corr: one_dim_extra_no_nugget}, with $n+1$ locations, where $n\ge2$. Suppose only the observation at $s_j$ is missing, for some $j\in\{2,\ldots,n\}$. If
\begin{equation*}
d(h,n+1)\le1,
\end{equation*}
then the target $s^*=s_j$ satisfies condition~(\ref{cond:C}) with respect to the remaining $n$ observed locations. Consequently, its KALE prediction MSE is lower than its zero-input-error value for all sufficiently small positive input-error variances.
\end{corollary}

\begin{proof}
    See Section \ref{Pf:cor2}
\end{proof}

Compared with Corollary~\ref{corr: one_dim_extra_no_nugget}, the missing-value result requires the additional condition $d(h,n+1)\le1$. Table~\ref{tab:values_of_c} shows that this condition holds whenever $h\ge1.15$, regardless of $n$. However, when $h=1$, the condition does not hold for any $n$.

Under the same grid conditions, both forecasting and missing-value imputation continue to exhibit local improvement in the presence of additive observation noise. Since condition~(\ref{cond:D}) is less restrictive than condition~(\ref{cond:C}), a sharper distance bound can be obtained in this setting.

\begin{corollary}[Forecasting and imputation in the presence of additive observation noise]
\label{corr: nugget_extension}
Fix a nugget-to-process variance ratio $\eta\ge0$, and define
\begin{equation*}
\widetilde d(h,n;\eta)
=
\sqrt{
\frac{[1-D_0(h,n)]d(h,n)^2+\eta/(2h^2)}
{1+\eta-D_0(h,n)}
}.
\end{equation*}
Under the respective grid and model assumptions, Corollaries~\ref{corr: one_dim_extra_no_nugget} and~\ref{corr: imputing_missing} extend to the nugget case with $d$ replaced by $\widetilde d$. Specifically, the sufficient conditions become
\begin{equation*}
s^*\ge s_n+\widetilde d(h,n;\eta)k
\qquad\text{and}\qquad
\widetilde d(h,n+1;\eta)\le1
\end{equation*}
for forecasting and missing-value imputation, respectively. In either case, the corresponding prediction target satisfies condition~(\ref{cond:D}).
\end{corollary}

\begin{proof}
    See Section \ref{Pf:cor3}.
\end{proof}

For fixed $h$ and $n$, $\widetilde d(h,n;\eta)$ is strictly decreasing in $\eta$, and $\lim_{\eta\to\infty}\widetilde d(h,n;\eta)=1/(\sqrt{2}h)$. Thus, the sufficient condition becomes easier to satisfy as the observation-noise variance increases, which agrees with Proposition~\ref{thm: kale_nugget}. Values of $\widetilde d(h,n;\eta)$ for $\eta=0.25$ and the same configurations of $h$ and $n$ used in Table~\ref{tab:values_of_c} are provided in Table~\ref{tab:c-values-eta-quarter}.

\begin{table}[!ht]
\centering
\caption{Values of $\widetilde d(h,n;\eta)$ for $\eta=0.25$,
with the same layout and rounding convention as
Table~\ref{tab:values_of_c}.}
\label{tab:c-values-eta-quarter}
\begin{tabular}{@{}ccccccc@{}}
\toprule
& \multicolumn{6}{c}{Training sample size $n$} \\
\cmidrule(lr){2-7}
$h$ & $2$ & $5$ & $10$ & $25$ & $100$ & $\infty$ \\
\midrule
0.90 & 1.64 & 2.17 & 2.18 & 2.18 & 2.18 & 2.18 \\
1.00 & 1.18 & 1.29 & 1.29 & 1.29 & 1.29 & 1.29 \\
1.10 & 0.93 & 0.96 & 0.96 & 0.96 & 0.96 & 0.96 \\
1.15 & 0.84 & 0.85 & 0.85 & 0.85 & 0.85 & 0.85 \\
1.50 & 0.51 & 0.51 & 0.51 & 0.51 & 0.51 & 0.51 \\
1.75 & 0.41 & 0.41 & 0.41 & 0.41 & 0.41 & 0.41 \\
2.00 & 0.35 & 0.35 & 0.35 & 0.35 & 0.35 & 0.35 \\
3.00 & 0.24 & 0.24 & 0.24 & 0.24 & 0.24 & 0.24 \\
\bottomrule
\end{tabular}
\end{table}

\begin{remark}[Irregularly spaced nominal locations]
Regular spacing is not essential for either the forecasting or missing-value imputation result. For distinct nominal time points, define the minimum separation by
\begin{equation*}
k=\min_{i\ne j}|s_i-s_j|,
\qquad
h=\frac{k}{\ell}.
\end{equation*}
For missing-value imputation, the minimum is taken over the complete set of nominal time points, including the missing target. Provided $h\ge0.9$ and the corresponding target-distance conditions in Corollaries~\ref{corr: one_dim_extra_no_nugget}--\ref{corr: nugget_extension} hold, the same local improvement conclusions follow whether or not additive observation noise is present.
\end{remark}


\subsection{Numerical example}

We use $n=25$ observations, fix $\ell=1$, and consider regular grids with scaled spacings $h=k/\ell\in\{0.5,0.9,1.5\}$. We forecast $q=1,2,3$ grid steps beyond the last observation and impute a central missing value on a grid of $26$ locations. We consider $\eta=0$ and $\eta=0.25$, with the latter representing additive measurement error. For each model and target, we compute the KALE prediction MSE from the closed-form covariances. The figures report the percentage reduction relative to zero Gaussian input-error variance at the same value of $\eta$, $100 \times\frac{M_{n,\eta}(0)-M_{n,\eta}(\sigma_u^2)}{M_{n,\eta}(0)}$.

\begin{figure}[!ht]
    \centering
        \includegraphics[width=10cm]{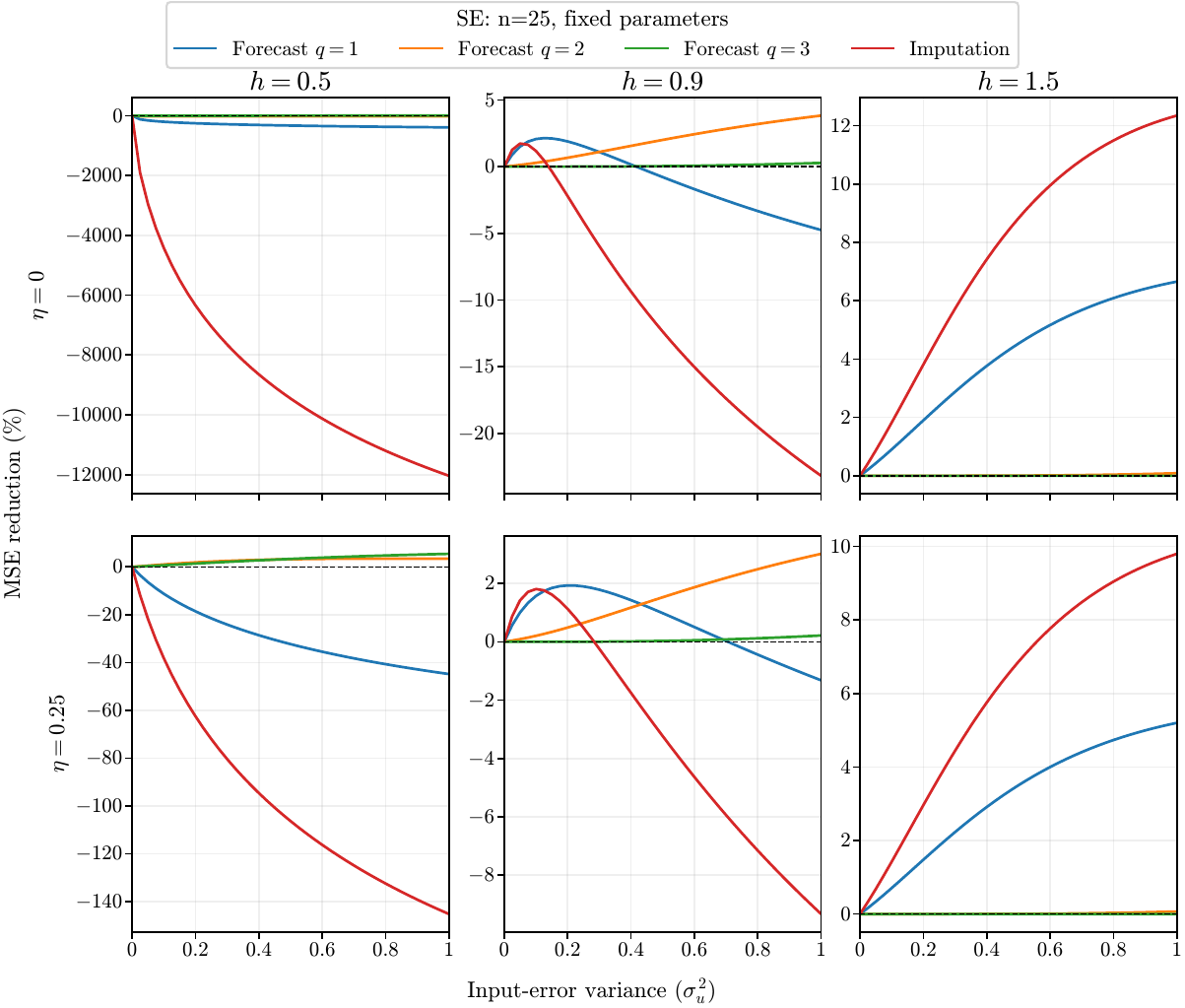}
\caption{Percentage reduction in KALE prediction MSE under SE--GE. Each curve uses its MSE at $\sigma_u^2=0$ as the baseline. Columns give the scaled grid spacing $h=k/\ell$, rows give the
nugget ratio $\eta$, and the dashed line marks zero reduction.}
\label{fig:se_mse}
\end{figure}

For SE--GE (Figure~\ref{fig:se_mse}), $h=0.5$ falls below the spacing requirement $h\ge0.9$ of corollaries \ref{corr: one_dim_extra_no_nugget}--\ref{corr: nugget_extension}. With measurement error, the MSE nevertheless decreases initially for the two- and three-step forecasts, which explains that local improvement can occur outside the range covered by the sufficient grid conditions.

At $h=0.9$, Tables~\ref{tab:values_of_c} and~\ref{tab:c-values-eta-quarter} give forecasting bounds of $6.86$ grid steps for $\eta=0$ and $2.18$ for $\eta=0.25$. The corresponding minimum integer horizons are $q=7$ and $q=3$. Both imputation bounds
exceed one, so among the targets shown the corollaries certify only the three-step forecast with measurement error. The figure also shows
an initial decrease in MSE for the shorter forecasts and imputation. At $h=1.5$, the forecasting and imputation bounds are below one at both values of $\eta$. The corollaries therefore guarantee local improvement for all four targets. Imputation gives the largest reduction over the plotted range.
\begin{figure}[!ht]
\centering
\includegraphics[width=10cm]{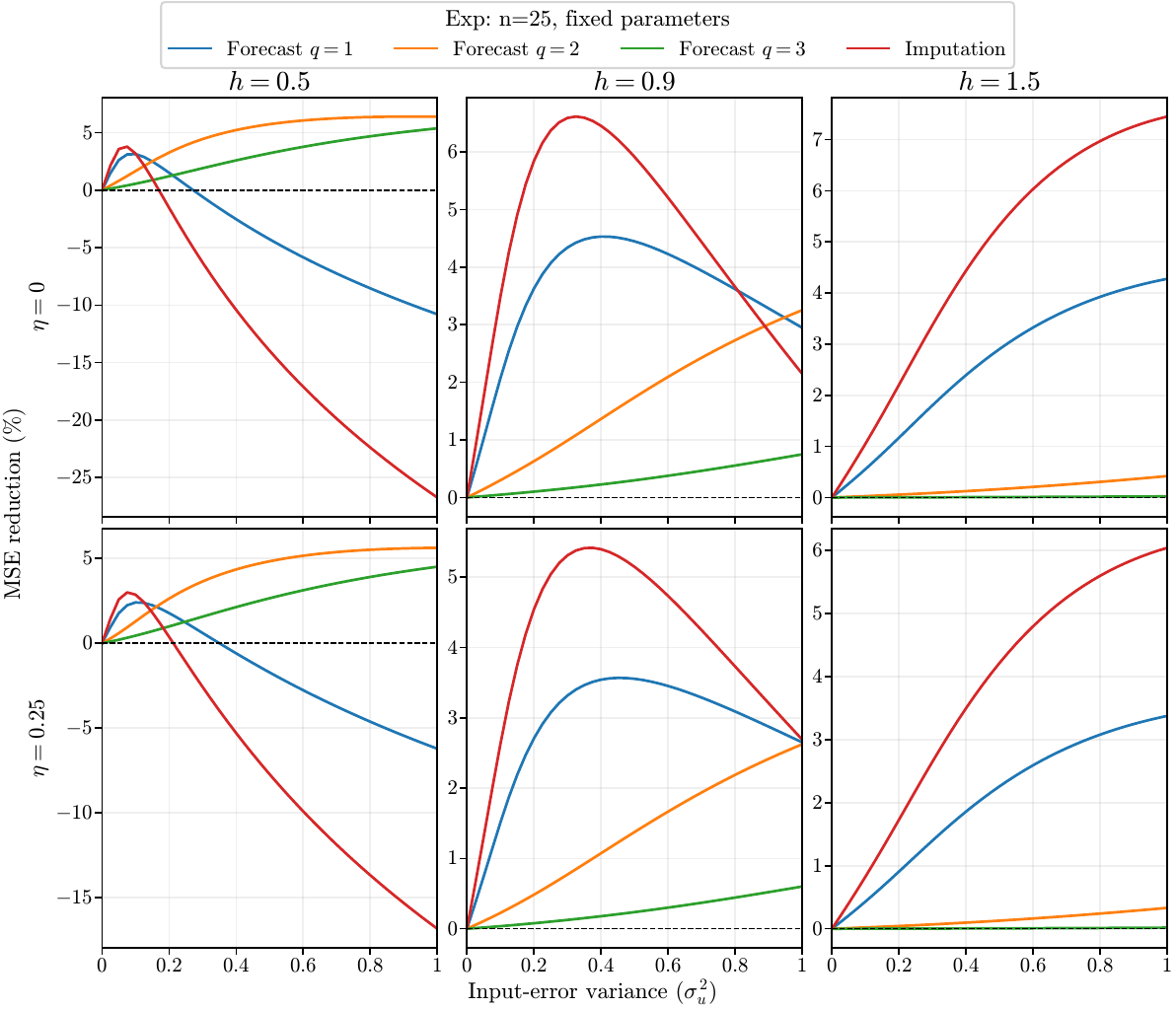}
\caption{Percentage reduction in KALE prediction MSE under Exp--GE,
with the same panel layout and baseline convention as
Figure~\ref{fig:se_mse}.}
\label{fig:exp_mse}
\end{figure}

For Exp--GE (Figure~\ref{fig:exp_mse}),
Proposition~\ref{thm:exponential_local_improvement} guarantees local improvement for all targets shown. At $h=0.5$, the MSEs for the one-step forecast and imputation eventually rise above their
zero-input-error baselines as $\sigma_u^2$ increases.

\section{Discussion}

We studied when Gaussian input error leads to local improvement in the KALE prediction MSE under fixed model parameters. Theorem~\ref{thm:exact} gives a first-order criterion for improvement as the input-error variance increases from zero. For the SE--GE model, we derived distance-based sufficient conditions with and without
additive observation noise and applied them to forecasting and imputation in sparsely sampled time series. For the one-dimensional Exp--GE model, we established local improvement for every finite set of distinct nominal training locations and every prediction target
outside that set.

The numerical examples show that the SE--GE prediction MSE can decrease initially in configurations outside the scope of the sufficient bounds for sparsely sampled time series. All Exp--GE configurations considered exhibit local improvement, as predicted by Proposition~\ref{thm:exponential_local_improvement}. The magnitude of the reduction varies with the grid spacing and prediction task. In some cases, the MSE rises above its zero-input-error value as the input-error variance increases further. In addition, our simulation study in Appendix~\ref{sec: simulation_estimation_improvement} shows that improvement can persist when the covariance parameters are estimated while the input-error variance is treated as known. However, gains obtained with known parameters are not retained in every configuration after estimation. Please see Appendix~\ref{sec: simulation_estimation_improvement} for more details.

Relevant future directions include developing necessary and sufficient distance-based conditions for local improvement of MSE under input error. Our results establish improvement for sufficiently small positive input-error variances, but do not identify an explicit improving range or an optimal variance. Obtaining such quantities requires control of the prediction MSE beyond its first-order behavior. Another direction is to establish a criterion that accounts for parameter estimation, whose effect on prediction risk requires a separate analysis \citep{zimmerman_cressie_1992, cervone2015loc_error}. 

Finally, our results do not imply that input error is generally beneficial or should be deliberately introduced. Rather, they explain a prediction phenomenon under fixed model parameters: observations collected at randomly perturbed inputs can yield a smaller KALE prediction MSE for a particular target than observations collected at the corresponding nominal inputs.

\section*{Declaration of generative AI use}

During the preparation of this work, the authors used ChatGPT (OpenAI) to assist with language editing and development of simulation code. The authors reviewed and revised the AI-assisted material and take full responsibility for the content of this article.


\newpage

\bibliographystyle{apalike}
\bibliography{references}

\newpage

\appendix

\section{Derivation of the closed forms for the SE-GE and Exp-GE smoothed covariances}

\label{derivation_closed_form}

\subsection{SE-GE model}
For the squared-exponential covariance, let $Z\sim N_p(\mathbf 0,\sigma^2\mathbf I_p)$ with $\sigma^2>0$. By stationarity,
\begin{align*}
c_{\sigma^2}(s,t)
&=\mathbb E[c(s+Z,t)]\\
&=\tau^2\mathbb E\left[
\exp\left\{-\frac{\|s-t+Z\|_2^2}{\ell^2}\right\}
\right].
\end{align*}
Since $(s-t+Z)/\sigma$ is normal with mean $(s-t)/\sigma$ and covariance $\mathbf I_p$, $\|s-t+Z\|_2^2/\sigma^2$ follows a noncentral $\chi^2$ distribution with $p$ degrees of freedom and noncentrality parameter $\lambda=\|s-t\|_2^2/\sigma^2$. Then using the moment-generating function of the noncentral $\chi^2$ distribution \citep{chi_mgf}, we get
\begin{equation*}
c_{\sigma^2}(s,t)
=\tau^2\left(1+\frac{2\sigma^2}{\ell^2}\right)^{-p/2}
\exp\left\{-\frac{\|s-t\|_2^2}{\ell^2+2\sigma^2}\right\}.
\end{equation*}

At $\sigma^2=0$, this expression reduces to $c(s,t)$. A more general expression allowing input-specific Gaussian covariance matrices is derived by \citet{dallaire2011approximate}.

\subsection{One dimensional Exp-GE model}
For the one-dimensional exponential covariance, let $Z\sim N(0,\sigma^2)$ with $\sigma^2>0$, and set $h=|s-t|$. By stationarity and the symmetry of $Z$,
\begin{align*}
c_{\sigma^2}(s,t)
&=\mathbb E[c(s+Z,t)]\\
&=\tau^2\mathbb E\left[e^{-|s-t+Z|/\ell}\right]
=\tau^2\mathbb E\left[e^{-|h+Z|/\ell}\right].
\end{align*}
Since $X=h+Z\sim N(h,\sigma^2)$, its absolute value $|X|$
has a folded normal distribution. Splitting the expectation
at $X=0$ and completing the square in each term gives the
Laplace transform of the folded normal distribution
\citep[Eq.~(25)]{gaussian_identity}. Evaluating it at $1/\ell$ yields
\begin{align*}
c_{\sigma^2}(s,t)
&=\tau^2\left\{
\mathbb E\left[e^{-X/\ell}\mathbf 1_{\{X\ge0\}}\right]
+\mathbb E\left[e^{X/\ell}\mathbf 1_{\{X<0\}}\right]
\right\}\\
&=\tau^2e^{\sigma^2/(2\ell^2)}
\Bigg[
e^{-h/\ell}
\Phi\left(\frac{h}{\sqrt{\sigma^2}}
-\frac{\sqrt{\sigma^2}}{\ell}\right)
+
e^{h/\ell}
\Phi\left(-\frac{h}{\sqrt{\sigma^2}}
-\frac{\sqrt{\sigma^2}}{\ell}\right)
\Bigg],
\end{align*}
where $\Phi$ is the standard normal cumulative distribution
function. At $\sigma^2=0$, $Z=0$ almost surely, so
$c_0(s,t)=\tau^2e^{-h/\ell}=c(s,t)$.



\section{Proof of Corollary~\ref{corr: one_dim_extra_no_nugget}}
\label{Pf:cor1}

For this proof, we write
\begin{equation*}
D_0=D_0(h,n),
\qquad
D_2=D_2(h,n),
\qquad
d=d(h,n).
\end{equation*}
Now, since $h\ge0.9$,
\begin{equation*}
0<D_0(h,n)
<
\frac{2e^{-h^2}}{1-e^{-3h^2}}
\le
\frac{2e^{-0.9^2}}{1-e^{-3(0.9)^2}}
<1.
\end{equation*}
Additionally, $4e^{-3h^2}\le4e^{-3(0.9)^2}<1$, so $D_2$ and $d$ are finite and strictly positive.

Set $r_i=(s^*-s_i)/\ell>0$ and define
\begin{equation*}
\label{eq: row_dom_rhs}
\mathcal R_i
=
\sum_{j\ne i}
e^{-H_{ij}}
\left|D_i+D_j-2H_{ij}\right|.
\end{equation*}

Fix an index $i$. At a separation of $m$ grid intervals from $s_i$, the possible indices are $j=i-m$ and $j=i+m$, which gives
\begin{equation*}
D_i+D_j-2H_{ij}
=
2r_i^2\pm2r_i mh-(mh)^2.
\end{equation*}
Extending both sums to $m=1,\ldots,n-1$ therefore yields
\begin{align*}
\mathcal R_i
&=
\sum_{m=1}^{i-1}
e^{-(mh)^2}
\left|2r_i^2+2r_i(mh)-(mh)^2\right|
+
\sum_{m=1}^{n-i}
e^{-(mh)^2}
\left|2r_i^2-2r_i(mh)-(mh)^2\right|\\
&\le
\sum_{m=1}^{n-1}
e^{-(mh)^2}
\Big[
\left|2r_i^2+2r_i(mh)-(mh)^2\right|
+
\left|2r_i^2-2r_i(mh)-(mh)^2\right|
\Big]\\
&=
2\sum_{m=1}^{n-1}
e^{-(mh)^2}
\max\Big\{
\left|2r_i^2-(mh)^2\right|,
2r_i(mh)
\Big\}\\
&<
2\sum_{m=1}^{n-1}
e^{-(mh)^2}
\Big[2r_i^2+(mh)^2\Big]\\
&=
4r_i^2\sum_{m=1}^{n-1}e^{-(mh)^2}
+
2\sum_{m=1}^{n-1}(mh)^2e^{-(mh)^2}.
\end{align*}
The equality involving the maximum follows from $|a+b|+|a-b|=2\max\{|a|,|b|\}$. The strict inequality follows because $r_i>0$ and $mh>0$ imply $\left|2r_i^2-(mh)^2\right|<2r_i^2+(mh)^2$, and $2r_i(mh)\le r_i^2+(mh)^2<2r_i^2+(mh)^2$.

For every positive integer $m$, $(m-1)(m-2)\ge0$, and hence $m^2\ge1+3(m-1)$. Moreover, $m^2\le4^{m-1}$. Applying these inequalities and the finite geometric-sum formula gives
\begin{equation}
\label{eq: one-dim-distance-bound}
\mathcal R_i
<
2D_0r_i^2+D_2.
\end{equation}

The target condition ensures
\begin{equation*}
r_i
=
\frac{s^*-s_i}{\ell}
\ge
\frac{s^*-s_n}{\ell}
\ge
h d.
\end{equation*}
Since $D_i=r_i^2$ and $1-D_0>0$, we have
$1+D_0D_2>1-D_0^2>0$. Consequently, the definition of $d$ gives
\begin{equation*}
h d
>
\frac{\sqrt{h^2+2+2\sqrt{2}h}-h}{2}
=
\frac{1}{\sqrt{2}}.
\end{equation*}
Thus, $r_i\ge hd(h,n)>1/\sqrt{2}$, and hence
$2D_i-1=2r_i^2-1>0$ for every $i$.

Now, define the function
\begin{equation*}
F(x)=\frac{2D_0x^2+D_2}{2x^2-1}
=
D_0+\frac{D_0+D_2}{2x^2-1},
\end{equation*}
which is positive and strictly decreasing for $x>1/\sqrt{2}$. The definition of $d$ gives
\begin{equation*}
2(1-D_0^2)h^2d(d+1)-(1+D_0D_2)
=
\sqrt{
(D_0+D_2)^2
+
2h^2(1-D_0^2)(1+D_0D_2)
}.
\end{equation*}
Squaring both sides and using $(1+D_0D_2)^2+(1-D_0^2)(D_2^2-1)=(D_0+D_2)^2$, we obtain
\begin{equation}
\label{eq:forecasting-boundary}
F(hd)F(hd+h)=1.
\end{equation}

For any two distinct indices $i$ and $j$, both target distances
are at least $r_n$, and at least one is at least $r_n+h$.
Therefore,
\begin{align*}
\frac{\mathcal R_i\mathcal R_j}
     {(2D_i-1)(2D_j-1)}
&<
\frac{2D_0r_i^2+D_2}{2r_i^2-1}
\frac{2D_0r_j^2+D_2}{2r_j^2-1}\\
&= F(r_i) F(r_j) \\
&\le
F(r_n) F(r_n+h)\\
&\le
F(hd) F(hd+h)\\
&=1.
\end{align*}
Together with the positive diagonal contributions, this establishes condition~(\ref{cond:C}).


\section{Proof of Corollary~\ref{corr: imputing_missing}}
\label{Pf:cor2}

Use the notation and bounds from Section~\ref{Pf:cor1}, with $n$ replaced by $n+1$ throughout. Since $d\le1$ and $h>0$, we have $hd\le h$ and $hd+h\le2h$. Also, because $F$ is positive and decreasing,
\begin{equation*}
F(h)\le F(hd),
\qquad
F(2h)\le F(hd+h).
\end{equation*}
Multiplying both inequalities and using \eqref{eq:forecasting-boundary}, we get
\begin{equation*}
F(h)F(2h)
\le
F(hd)F(hd+h)
=
1.
\end{equation*}
Moreover, $D_0\ge2e^{-h^2}$ and $D_2\ge2h^2e^{-h^2}$. As a result,
\begin{equation}
\label{eq:imputation-product-bound}
1
\ge
F(h)F(2h)
\ge
\frac{108h^4e^{-2h^2}}{(2h^2-1)(8h^2-1)}.
\end{equation}
We must therefore have $h>\sqrt{5}/2>1$. To see this, suppose $h\le\sqrt{5}/2$. Since the final expression in \eqref{eq:imputation-product-bound} is strictly decreasing for $h\ge0.9$, we would have $F(h)F(2h)\ge(25/2)e^{-5/2}>1$, which is a contradiction.

For each observed index $i\ne j$, write $r_i=|s_j-s_i|/\ell=|i-j|h$, and compute $\mathcal R_i$ using only the observed locations. Taking absolute target distances merely interchanges the two terms in the paired row sum from Section~\ref{Pf:cor1}. Thus, \eqref{eq: one-dim-distance-bound} remains applicable. Also, $r_i\ge h>1$, so $2D_i-1>0$ for every observed index.

For the immediate neighbors, $r_i=h$. The paired contributions at $m=1$ and $m=2$ are $4h^2e^{-h^2}$ and $8h^2e^{-4h^2}$, whereas the bounds in Section~\ref{Pf:cor1} are $6h^2e^{-h^2}$ and
$12h^2e^{-4h^2}$. Removing the missing observation further subtracts $h^2e^{-h^2}$. These corrections give the first row bound below. The inequality remains strict because the extended sum includes positive contributions from outside the finite grid. For every other observed index, $r_i\ge2h$, so the decreasing property of $F$ gives the second row bound.

The two row bounds used below increase with $n$, and their limits decrease with $h\ge1$. Since $h>\sqrt{5}/2$, we evaluate these limiting bounds at $h=\sqrt{5}/2$.

For the immediate neighbors, $|i-j|=1$, we obtain
\begin{align*}
\frac{\mathcal R_i}{2D_i-1}
&<
\frac{
2h^2D_0+D_2-3h^2e^{-h^2}-4h^2e^{-4h^2}
}{2h^2-1}\\
&\le
\lim_{n\to\infty}
\frac{
2h^2D_0(h,n+1)+D_2(h,n+1)
-3h^2e^{-h^2}-4h^2e^{-4h^2}
}{2h^2-1}\\
&=
\frac{h^2e^{-h^2}}{2h^2-1}
\left[
\frac{4}{1-e^{-3h^2}}
+
\frac{2}{1-4e^{-3h^2}}
-3-4e^{-3h^2}
\right]\\
&\le
\frac{5e^{-5/4}}{6}
\left[
\frac{4}{1-e^{-15/4}}
+
\frac{2}{1-4e^{-15/4}}
-3-4e^{-15/4}
\right]\\
&<0.77<1.
\end{align*}

For the remaining observed indices, $|i-j|\ge2$, we similarly get
\begin{align*}
\frac{\mathcal R_i}{2D_i-1}
&<
F(2h)
=
\frac{8h^2D_0+D_2}{8h^2-1}\\
&\le
\lim_{n\to\infty}
\frac{
8h^2D_0(h,n+1)+D_2(h,n+1)
}{8h^2-1}\\
&=
\frac{h^2e^{-h^2}}{8h^2-1}
\left[
\frac{16}{1-e^{-3h^2}}
+
\frac{2}{1-4e^{-3h^2}}
\right]\\
&\le
\frac{5e^{-5/4}}{36}
\left[
\frac{16}{1-e^{-15/4}}
+
\frac{2}{1-4e^{-15/4}}
\right]\\
&<0.75<1.
\end{align*}
Hence,
\begin{equation*}
\mathcal R_i<2D_i-1
\end{equation*}
for every observed index $i$. Hence, for every pair of distinct observed indices $i$ and $l$,
\begin{equation*}
\mathcal R_i\mathcal R_l
<
(2D_i-1)(2D_l-1).
\end{equation*}
Together with the positive diagonal contributions, this establishes condition~(\ref{cond:C}).


\section{Proof of Corollary~\ref{corr: nugget_extension}}
\label{Pf:cor3}

Use the notation from Sections~\ref{Pf:cor1}--\ref{Pf:cor2},
with sample-size argument $n$ for forecasting and $n+1$ for
imputation. Define
\begin{equation*}
\lambda=\frac{1+\eta-D_0}{1-D_0}\ge1.
\end{equation*}
The definition of $\widetilde d$ gives
\begin{equation*}
2h^2\widetilde d^2-1
=
\frac{2h^2d^2-1}{\lambda}>0,
\qquad
\lambda\widetilde d\ge d.
\end{equation*}
Consequently,
$2h^2(\widetilde d+1)^2-1
\ge[2h^2(d+1)^2-1]/\lambda$.
Using the expression for $F$ in Section~\ref{Pf:cor1}, we obtain
\begin{align*}
F(h\widetilde d)F(h\widetilde d+h)
&\le
[\lambda F(hd)-(\lambda-1)D_0]
[\lambda F(hd+h)-(\lambda-1)D_0]\\
&\le
[\lambda-(\lambda-1)D_0]^2\\
&=(1+\eta)^2.
\end{align*}
The second inequality follows from
\eqref{eq:forecasting-boundary} and
$F(hd)+F(hd+h)\ge2$.

For forecasting, $r_n\ge h\widetilde d>1/\sqrt{2}$.
The same row-ratio argument as in Section~\ref{Pf:cor1}
therefore gives
\begin{equation*}
\frac{\mathcal R_i\mathcal R_l}
     {(2D_i-1)(2D_l-1)}
<
F(h\widetilde d)F(h\widetilde d+h)
\le(1+\eta)^2,
\qquad i\ne l.
\end{equation*}

For imputation, use the corrected neighbor-row bound from
Section~\ref{Pf:cor2}. This bound remains valid for $h\ge0.9$.
To compare it with $\sqrt{F(h)F(2h)}$, write
$q=e^{-3h^2}<0.09$. The geometric-sum formulas give
\begin{equation*}
D_2
\le
\frac{2h^2e^{-h^2}}{1-4q}
<
h^2e^{-h^2}(3+4q).
\end{equation*}
Over these ranges, the fraction below increases with both
$D_0$ and $D_2$. Replacing them by their limits as $n\to\infty$
gives
\begin{align*}
&
\frac{
2h^2D_0+D_2-3h^2e^{-h^2}-4h^2e^{-4h^2}
}{
\sqrt{(2h^2D_0+D_2)(8h^2D_0+D_2)}
}\\
&\qquad\le
\frac{3-7q+8q^2-16q^3}
     {6\sqrt{(1-3q)(3-11q)}}\\
&\qquad<
0.334
<
\sqrt{\frac{31}{274}}
\le
\sqrt{\frac{2h^2-1}{8h^2-1}}.
\end{align*}
The expression in $q$ is increasing on $[0,0.09]$, as follows
by differentiation, and its value at $q=0.09$ is smaller than
$0.334$. The last inequality uses $h\ge0.9$.

Thus, the corrected neighbor-row bound is smaller than
$\sqrt{F(h)F(2h)}$. For the other observed indices,
Section~\ref{Pf:cor2} gives the bound
$F(2h)\le\sqrt{F(h)F(2h)}$.
Hence, when $\widetilde d\le1$, every observed index satisfies
\begin{equation*}
\frac{\mathcal R_i}{2D_i-1}
<
\sqrt{F(h)F(2h)}
\le
\sqrt{F(h\widetilde d)F(h\widetilde d+h)}
\le1+\eta.
\end{equation*}
The middle inequality uses the decreasing property of $F$.

Finally, the diagonal contributions are positive:
$r_i\ge h\widetilde d>1/\sqrt{2}$ for forecasting and
$|r_i|\ge h\ge0.9>1/\sqrt{2}$ for imputation.
Thus, condition~(\ref{cond:D}) holds in both cases.
The conclusions follow from
Proposition~\ref{thm: kale_nugget} for $\eta>0$ and
Proposition~\ref{thm: kale_no_nugget} for $\eta=0$.

\section{Properties of \texorpdfstring{$\tilde d(h,n,\eta)$}{c-tilde(h,n,eta)}}

For fixed $h$ and $n$,
\begin{equation*}
\frac{\partial\widetilde d(h,n;\eta)}{\partial\eta}
=
-\frac{ [1-D_0(h,n)] \left[d(h,n)^2-\dfrac{1}{2h^2}\right] }{
2\widetilde d(h,n;\eta)[1+\eta-D_0(h,n)]^2
} <0.    
\end{equation*}

Thus, $\widetilde d(h,n;\eta)$ is strictly decreasing in $\eta$. Additionally, $\widetilde d(h,n;\eta)$ equals $d(h,n)$ at $\eta=0$, and has limit 

\begin{equation*}
\displaystyle \lim_{\eta\to\infty}\widetilde d(h,n;\eta)=\lim_{\eta\to\infty}\sqrt{\frac{[1-D_0(h,n)]d(h,n)^2/\eta+1/(2h^2)}{[1-D_0(h,n)]/\eta+1}}=\frac{1}{\sqrt{2}h}.
\end{equation*}
\section{Simulation study: Local improvemnt of KALE MSE under estimated parameters}

\label{sec: simulation_estimation_improvement}

\subsection{Simulation setup}

We investigate whether the improvement in prediction MSE found with known covariance parameters persists when these parameters are estimated. We conduct the experiment separately for the SE--GE and Exp--GE models. The generating parameters are $\ell=\tau^2=1$, with $n=25$ training observations and $R=500$ replications for each setting. We consider $h=k/\ell\in\{0.5,0.9,1.5\}$,
$\eta\in\{0,0.25\}$, and the input-error variances
\begin{equation*}
\mathcal V
=
\{\sigma_{u,j}^2:j=0,\ldots,J\}
=
\{0,0.025,0.05,0.10,0.25,0.50,1\},
\qquad J=6.
\end{equation*}
This set covers the range from zero to one, with more values near zero. For forecasting, the training times are $s_i=(i-1)k$, $i\in[n]$, and the targets are $s^*=s_n+qk$, $q\in\{1,2,3\}$. For imputation, we use the grid $0,k,\ldots,nk$ and remove the observation at $s^*=\lfloor n/2\rfloor k$. This selects the earlier middle time when there are two middle points and leaves $n$ training observations.

In replication $r\in[R]$, the latent signal $g_r$ is an independent realization of a zero-mean GP with the generating covariance. At each nominal grid time, we generate an independent standard-normal input-error draw and an independent observation-error draw. Writing these as $z_{ir}\sim N(0,1)$ and $\epsilon_{ir}\sim N(0,\eta\tau^2)$ at a training time $s_i$, we set $u_{ij}^{(r)}=\sigma_{u,j}z_{ir}$ and construct \begin{equation*}
\mathcal D_j^{(r)}
=
\left\{
\left(s_i,g_r(s_i+u_{ij}^{(r)})+\epsilon_{ir}\right)
\right\}_{i=1}^n,
\qquad j=0,\ldots,J.
\end{equation*}
Conditional on the input errors, signal values at all nominal times, perturbed times, and targets are generated jointly using the original covariance evaluated at the actual times. The same GP realization and the same input-error and observation-error draws are shared across variance levels within each replication. Draws at shared nominal times are also shared between the forecasting and imputation designs. 

Each dataset is used by two predictors. The first uses the generating covariance parameters, while the second estimates $\ell$, $\tau^2$, and $\eta$ separately for each dataset. Both use the generating covariance family and treat $\sigma_{u,j}^2$ as known. At zero input-error variance, each predictor reduces to exact-input kriging and provides its own baseline. For each task, both predictors and all variance levels use the same latent target $g_r(s^*)$, which is excluded from that task's training sample. One forecasting fit predicts all three future targets, while imputation uses a separate training sample and fit.

For parameter estimation, we use the Gaussian likelihood for exact-input kriging and the Gaussian quasi-likelihood for KALE \citep{cervone2015loc_error}. We estimate the nugget ratio in both generating noise settings. We restrict $\ell$ to $[0.2,3]$ and $\eta$ to $[0,2]$, and profile out $\tau^2$. We evaluate a $5\times5$ initial grid and refine the four best candidates using L-BFGS \citep{liu1989limited,wenger2022preconditioning},
with at most $100$ iterations per starting point. The boundary
$\eta=0$ is optimized separately. For each training sample,
the exact-input fit is reused at zero input-error variance.

For each design and target, let
$\widehat g_{j,r}^{\mathrm{known}}(s^*)$ and
$\widehat g_{j,r}^{\mathrm{fit}}(s^*)$ be the predictions using known and estimated covariance parameters, respectively. Their squared errors are
\begin{equation*}
L_{j,r}^{a}
=
\left[
\widehat g_{j,r}^{a}(s^*)-g_r(s^*)
\right]^2,
\qquad
a\in\{\mathrm{known},\mathrm{fit}\}.
\end{equation*}
The known-parameter MSE,
$M_j^{\mathrm{known}}=M_{n,\eta}(\sigma_{u,j}^2)$, is calculated directly from the covariance matrices. For the fitted predictor, we use the known-parameter squared error as a control variate and estimate its MSE by
\begin{equation*}
\widehat M_j^{\mathrm{fit}}
=
M_j^{\mathrm{known}}
+
\frac1R\sum_{r=1}^R
\left(
L_{j,r}^{\mathrm{fit}}
-
L_{j,r}^{\mathrm{known}}
\right).
\end{equation*}
We report the percentage reductions
\begin{equation*}
100 \times \left[
1-\frac{M_j^{\mathrm{known}}}{M_0^{\mathrm{known}}}
\right]
\qquad\text{and}\qquad
100\times \left[
1-\frac{\widehat M_j^{\mathrm{fit}}}
        {\widehat M_0^{\mathrm{fit}}}
\right].
\end{equation*}
Positive values indicate improvement relative to the corresponding zero-input-error baseline.

We assess Monte Carlo uncertainty using $4000$ paired bootstrap samples of the $R$ replications. Within each task
and setting, the same resampled replication indices are used for the baseline and every input-error variance. For each bootstrap sample, we recompute the fitted MSE estimates and their percentage reductions from the stored squared prediction errors. The simulation study was implemented in \texttt{Python} using \texttt{PyTorch}, with GPU acceleration through Apple's Metal Performance Shaders (\texttt{MPS}) backend.


\begin{figure}[!ht]
    \centering
        \includegraphics[width=14cm]{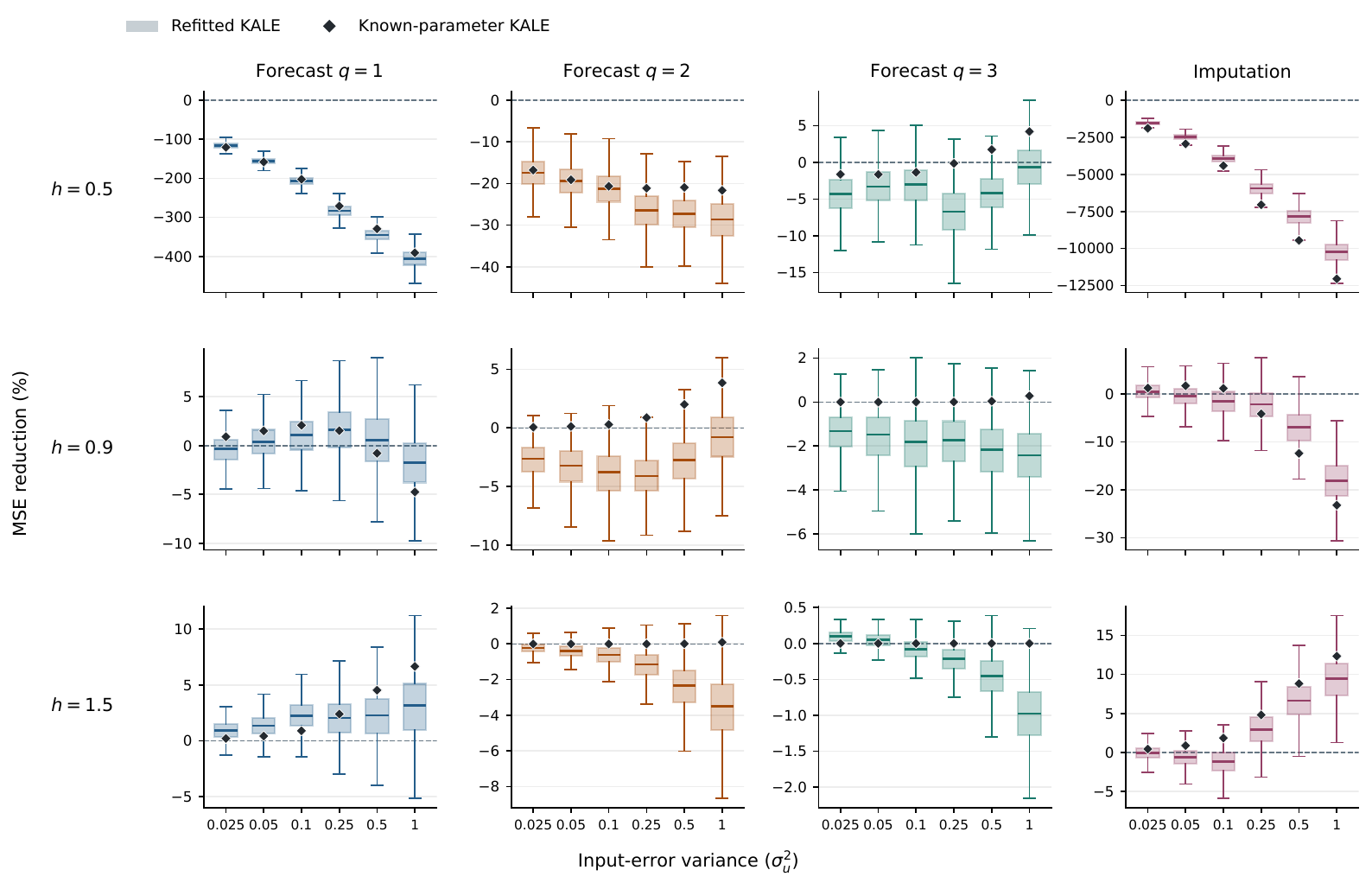}
\caption{Percentage reduction in KALE prediction MSE under
SE--GE with generating nugget ratio $\eta=0$. Rows correspond to scaled grid spacings $h=0.5,0.9,1.5$,
and columns to forecasting at $q=1,2,3$ and missing-value
imputation. Boxes summarize the paired bootstrap distributions of the estimated reductions after fitting the covariance parameters; black diamonds show the exact reductions with known parameters. Each predictor uses its own zero-input-error baseline, and positive values indicate improvement.}
\label{fig:se_est_eta0}
\end{figure}

\subsection{Simulation results}

\begin{figure}[!ht]
    \centering
        \includegraphics[width=14cm]{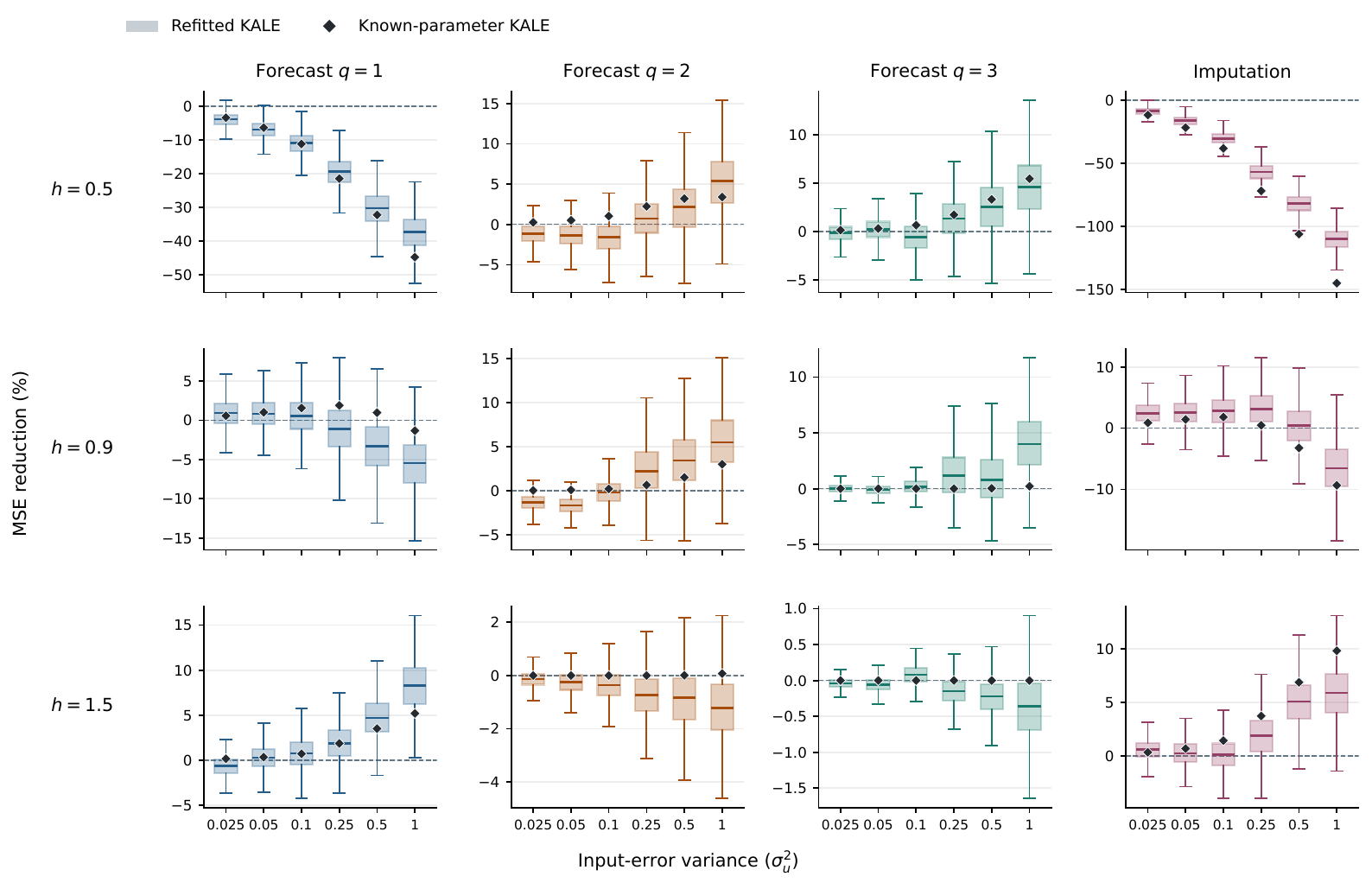}
\caption{Percentage reduction in KALE prediction MSE under SE--GE with generating nugget ratio $\eta=0.25$, with the same panel layout, baseline convention, and plotting conventions as Figure~\ref{fig:se_est_eta0}.}
\label{fig:se_est_eta025}
\end{figure}

We summarize the bootstrap distribution of the percent reduction of MSE from the simulation studies and the exact reductions in Figure~\ref{fig:se_est_eta0}--\ref{fig:se_est_eta025}. The simulation results show that improvement in prediction MSE can persist after covariance parameter estimation, but its occurrence depends on the covariance model, sampling spacing, observation noise, and prediction task. The fitted reductions sometimes follow the known-parameter results closely, while other settings show substantial differences in their size or direction. Several bootstrap distributions also extend across zero, particularly when the known-parameter reductions are small.

For SE--GE with $\eta=0$ (Figure~\ref{fig:se_est_eta0}), the clearest deterioration occurs at $h=0.5$. One-step and two-step forecasting and imputation show negative reductions throughout the tested variance grid.
For three-step forecasting, the known-parameter reduction becomes positive at larger input-error variances, but the fitted reductions remain centered below zero. At $h=0.9$, the known-parameter gains for two-step forecasting are also not retained in the centers of the bootstrap distributions. At $h=1.5$, positive fitted reductions are observed for one-step forecasting and, at larger input-error variances, imputation. The known-parameter gains for the longer forecasting horizons are much smaller, and the fitted reductions are often negative.

In the presence of observation noises, these patterns (Figure~\ref{fig:se_est_eta025}) change. At $h=0.5$, two-step and three-step forecasting show positive fitted reductions at larger input-error variances, although one-step forecasting and imputation still deteriorate. At $h=0.9$, the fitted reductions for the longer forecasting horizons also become positive as the input-error variance increases. Some of these percentage reductions exceed their known-parameter counterparts. At $h=1.5$, one-step forecasting and imputation again retain positive reductions at larger variances. The much larger percentage losses in the noiseless SE imputation setting partly reflect its very small zero-input-error baseline MSE. Consequently, the sizes of the percentage reductions across noise settings should be interpreted as relative changes from their respective baselines.
\begin{figure}[!ht]
    \centering
        \includegraphics[width=14cm]{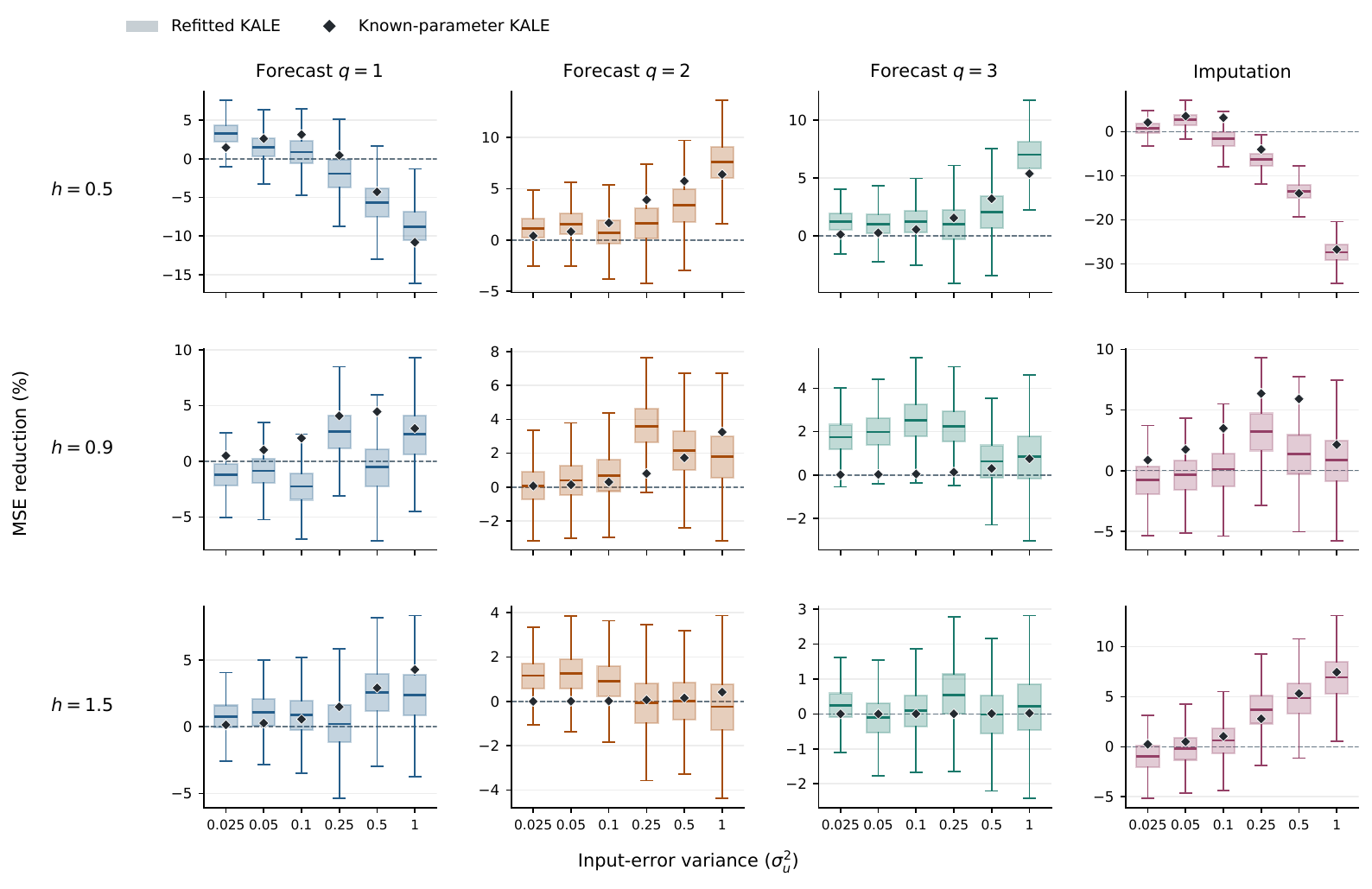}
\caption{Percentage reduction in KALE prediction MSE under Exp--GE with generating nugget ratio $\eta=0$,
with the same panel layout, baseline convention, and plotting conventions as Figure~\ref{fig:se_est_eta0}.}
\label{fig:exp_est_eta0}
\end{figure}

For Exp--GE with $\eta=0$
(Figure~\ref{fig:exp_est_eta0}), improvement after fitting is
observed in several forecasting configurations. At $h=0.5$,
the fitted reductions for two-step and three-step forecasting
remain centered above zero and increase at the largest tested
variance. In contrast, the gains for one-step forecasting
and imputation at small input-error variances give way to
deterioration as the variance increases. This change is
consistent with the local nature of the theoretical result.
At $h=0.9$, three-step forecasting shows positive fitted
reductions even though the known-parameter reductions are
small. Imputation at $h=1.5$ also shows clear positive
reductions at larger input-error variances.

\begin{figure}[!ht]
    \centering
        \includegraphics[width=14cm]{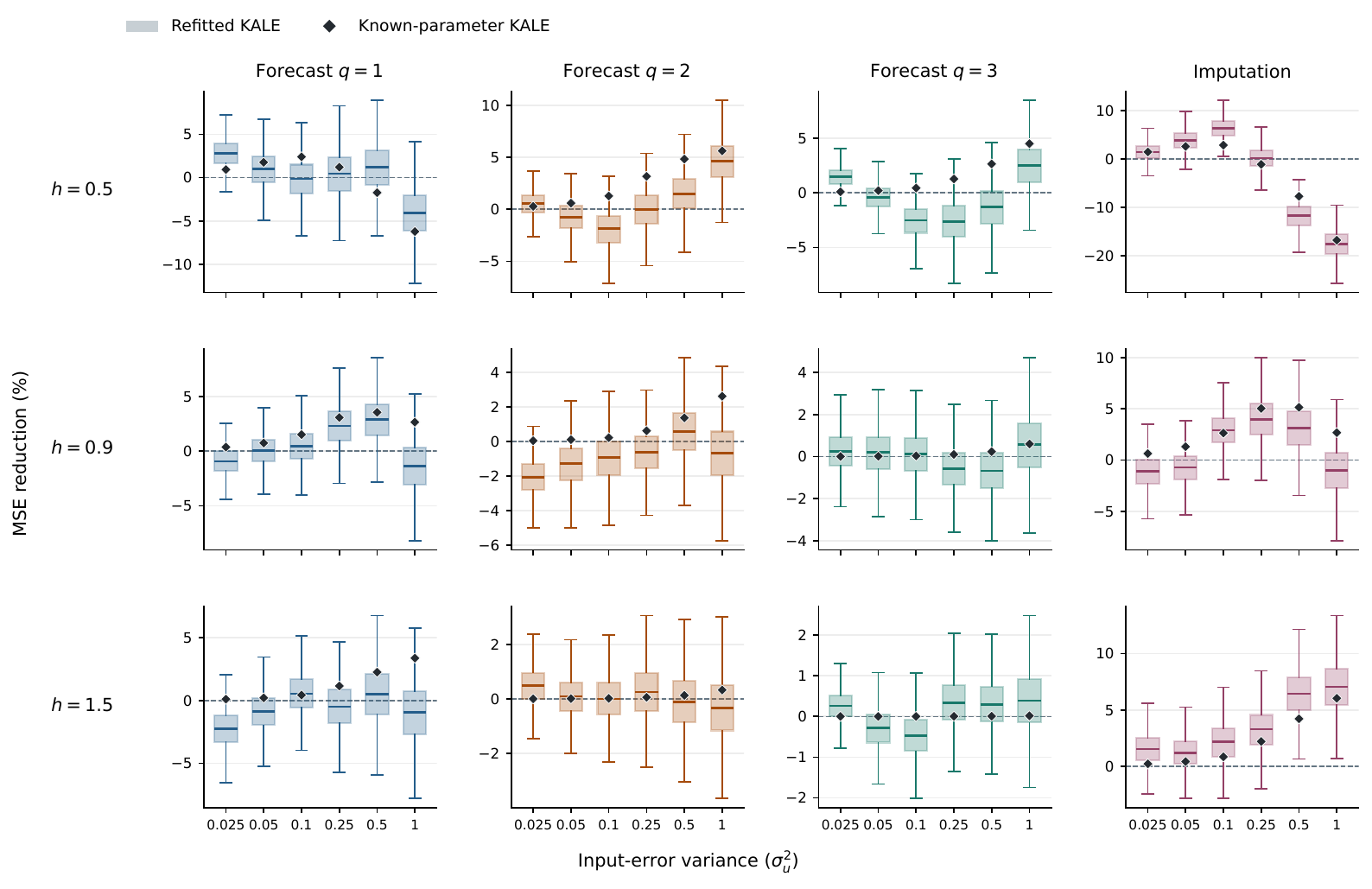}
\caption{Percentage reduction in KALE prediction MSE under
Exp--GE with generating nugget ratio $\eta=0.25$,
with the same panel layout, baseline convention, and
plotting conventions as Figure~\ref{fig:se_est_eta0}.}
\label{fig:exp_est_eta025}
\end{figure}

The Exp--GE results with $\eta=0.25$
(Figure~\ref{fig:exp_est_eta025}) show that observation noise
does not uniformly preserve the known-parameter gains.
For example, at $h=0.5$, the fitted three-step forecasting
reductions are centered below zero at several intermediate
variances, although the known-parameter reductions are
positive. At $h=1.5$, one-step forecasting also has mostly
negative fitted reductions despite positive known-parameter
reductions. Nevertheless, imputation at this spacing retains
positive reductions, particularly at larger input-error
variances. Thus, imputation at the widest spacing provides
a common example of retained improvement across both
covariance models and noise settings.

To relate these comparisons to the theoretical result,
let $P_{\sigma_u^2}$ and $\widehat P_{\sigma_u^2}$ denote
the KALE predictions using the generating and estimated
covariance parameters, respectively. Define the prediction
MSE after parameter estimation by
\begin{equation*}
R(\sigma_u^2)
=
\mathbb E\left[
\left(\widehat P_{\sigma_u^2}-g(s^*)\right)^2
\right].
\end{equation*}
Expanding the square, we get
\begin{equation}
\label{eq:mse_decomposition}
\begin{aligned}
R(\sigma_u^2)-R(0)
&=
\underbrace{
M_{n,\eta}(\sigma_u^2)-M_{n,\eta}(0)
}_{\textnormal{change with known parameters}}+
\underbrace{
\Delta_{\mathrm{est}}(\sigma_u^2)-\Delta_{\mathrm{est}}(0)
}_{\textnormal{change due to parameter estimation}},
\end{aligned}
\end{equation}
where
\begin{align*}
\Delta_{\mathrm{est}}(\sigma_u^2)
&=
\mathbb E\left[
\left(\widehat P_{\sigma_u^2}-P_{\sigma_u^2}\right)^2
\right]+
2\mathbb E\left[
\left(\widehat P_{\sigma_u^2}-P_{\sigma_u^2}\right)
\left(P_{\sigma_u^2}-g(s^*)\right)
\right].
\end{align*}
The cross-product term need not vanish, since the fitted
prediction depends on parameter estimates obtained from
the same observations. The paired differences in squared
errors used in the simulation estimate
$\Delta_{\mathrm{est}}(\sigma_u^2)$.

Theorem~\ref{thm:exact} determines the first-order behavior
of the known-parameter term, but does not control the change
in the contribution from parameter estimation. In particular, we have $R(\sigma_u^2)<R(0)$ if and only if $\Delta_{\mathrm{est}}(\sigma_u^2)-\Delta_{\mathrm{est}}(0)
<
M_{n,\eta}(0)-M_{n,\eta}(\sigma_u^2)$. Thus, a known-parameter improvement can disappear if the contribution from estimation increases enough to offset it. Conversely, a decrease in this contribution can strengthen the improvement after fitting. Since each predictor uses
its own zero-input-error baseline, a larger percentage
reduction after fitting does not by itself imply a smaller
absolute prediction MSE.

\end{document}